\documentclass[11pt, a4paper, logo, copyright]{googledeepmind}

\pdftrailerid{redacted}

\makeatletter
\renewcommand\bibentry[1]{\nocite{#1}{\frenchspacing\@nameuse{BR@r@#1\@extra@b@citeb}}}
\makeatother

\usepackage{kantlipsum, lipsum}
\usepackage{dsfont}
\usepackage{gdm-colors}
\usepackage[utf8]{inputenc}   
\usepackage{newunicodechar}   
\usepackage{amssymb}          
\usepackage{wasysym,marvosym}
\usepackage{ulem}

\usepackage{tikz}
\usetikzlibrary{shapes.geometric, arrows.meta, positioning}
\usepackage{amsmath}
\usepackage{algorithm}
\usepackage{algpseudocode}
\usepackage{listings}
\usepackage{enumitem} 
\usepackage{lipsum}
\usepackage[most]{tcolorbox}
\definecolor{thinkcolor}{RGB}{227,196,144}
\definecolor{observecolor}{RGB}{153,201,227}
\definecolor{explorecolor}{RGB}{178,217,200}

\tcbset{
    common/.style={
		enhanced,
		arc=0mm,
		fonttitle=\large\bfseries,
		coltitle=black,
		attach boxed title to top left={xshift=0mm,
										yshift=-0.50mm},
		boxed title style={
			skin=enhancedfirst jigsaw,
			size=small,
			arc=5mm,
			bottom=0mm,
			left=8mm,
			right=18mm,
			top=1mm},
			boxrule=0pt,
			frame hidden},
    thinkstyle/.style={
		common,
		colbacktitle=thinkcolor,
		colframe=thinkcolor,
		colback=thinkcolor!40,
		borderline north={4pt}{0pt}{thinkcolor}},
	observestyle/.style={
		common,
		colbacktitle=observecolor,
		colframe=observecolor,
		colback=observecolor!40,
		borderline north={4pt}{0pt}{observecolor}}
}

\newtcolorbox{think}{thinkstyle,title=Prompt Template}
\newtcolorbox{observe}{observestyle,title=observe}
\newtcolorbox{custom}[2][gray]{
	common,
	title=#2,
	colbacktitle=#1,
	colframe=#1,
	colback=#1!40,
	borderline north={4pt}{0pt}{#1}}
\usepackage{array, multirow, tabularx, booktabs, makecell}

\definecolor{interest_colframe}{rgb}{0.8, 0.878, 0.871}
\definecolor{interest_colback}{rgb}{0.918, 0.953, 0.949}

\definecolor{tag_colframe}{rgb}{0.965, 0.898, 0.847}
\definecolor{tag_colback}{rgb}{0.988, 0.961, 0.941}

\definecolor{exp_colback}{rgb}{0.949, 0.965, 0.980}
\definecolor{exp_colframe}{rgb}{0.878, 0.922, 0.965}

\tcbset{
    promptbox/.code args={#1/#2}{
        \tcbset{
            enhanced,
            arc=0mm,
            colframe=#1, 
            colback=#2, 
            coltitle=black,
            fonttitle=\large\bfseries,
            attach boxed title to top left={xshift=0mm, yshift=-1.0mm},
            boxed title style={
                skin=enhancedfirst jigsaw,
                size=small,
                arc=3mm,
                bottom=0mm,
                left=8mm,
                right=8mm,
                top=1mm,
                colback=#1
            },
            boxrule=0pt,
            frame hidden,
            borderline north={4pt}{0pt}{#1},
        }
    }
}

\newcommand{\assignmentQuestionName}{Question} 

\usepackage{caption}
\usepackage{booktabs}
\usepackage{arydshln}
\usepackage{dashrule}

\usepackage[authoryear, sort&compress, round]{natbib}

\usepackage{bbding}
\usepackage[T1]{fontenc}    
\usepackage{hyperref}       
\usepackage{url}            
\usepackage{booktabs}       
\usepackage{nicefrac}       
\usepackage{microtype}      
\usepackage{amsmath}
\usepackage{graphicx}
\usepackage{multicol}
\usepackage[nameinlink]{cleveref}
\usepackage{bbm}
\usepackage{multirow}
\usepackage{soul}
\usepackage{float}
\usepackage{wrapfig}
\usepackage{blindtext}
\usepackage{tablefootnote}
\usepackage{amsfonts}
\usepackage[flushleft]{threeparttable}
\usepackage{colortbl}
\usepackage{mathtools,amssymb}
\usepackage{bm}
\usepackage{makecell}
\usepackage{caption}
\usepackage{capt-of}
\usepackage{array}
\usepackage{calc}      
\usepackage{caption}   
\usepackage{subcaption}  
\usepackage[bottom]{footmisc}
\usepackage{fontawesome}
\usepackage{tabularx}        

\newcommand{\correct}{\textcolor{green}{\ding{51}}}
\newcommand{\wrong}{\textcolor{red}{\ding{55}}}

\graphicspath{{Figures/}}

\title{RecGPT Technical Report}

\renewcommand{\today}{}

\author{\large RecGPT Team}

\begin{abstract}
Recommender systems are among the most impactful applications of artificial intelligence, serving as critical infrastructure connecting users, merchants, and platforms.
However, most current industrial systems remain heavily reliant on historical co-occurrence patterns and log-fitting objectives—i.e., optimizing for past user interactions without explicitly modeling user intent. This log-fitting approach often leads to overfitting to narrow historical preferences, failing to capture users’ evolving and latent interests. As a result, it reinforces filter bubbles and long-tail phenomena, ultimately harming user experience and threatening the sustainability of the whole recommendation ecosystem.

To address these challenges, we rethink the overall design paradigm of recommender systems and propose RecGPT, a next-generation framework that places user intent at the center of the recommendation pipeline. 
By integrating large language models (LLMs) into key stages of user interest mining, item retrieval, and explanation generation, RecGPT transforms log-fitting recommendation into an intent-centric process. 
To effectively align general-purpose LLMs to the above domain-specific recommendation tasks at scale, RecGPT incorporates a multi-stage training paradigm, which integrates reasoning-enhanced pre-alignment and self-training evolution, guided by a Human-LLM cooperative judge system.
Currently, RecGPT has been fully deployed on the Taobao App. Online experiments demonstrate that RecGPT achieves consistent performance gains across stakeholders: users benefit from increased content diversity and satisfaction (\textit{e.g.}, CICD +6.96\%, DT +4.82\%), merchants and the platform gain greater exposure and conversions (\textit{e.g.}, CTR +6.33\%, IPV +9.47\%, DCAU +3.72\%). These comprehensive improvement results across all stakeholders validates that LLM-driven, intent-centric design can foster a more sustainable and mutually beneficial recommendation ecosystem.
\end{abstract}

\begin{document}
\maketitle

\vspace{-1em}
\begin{figure}[h!]
    \centering
    \includegraphics[width=0.9\textwidth]{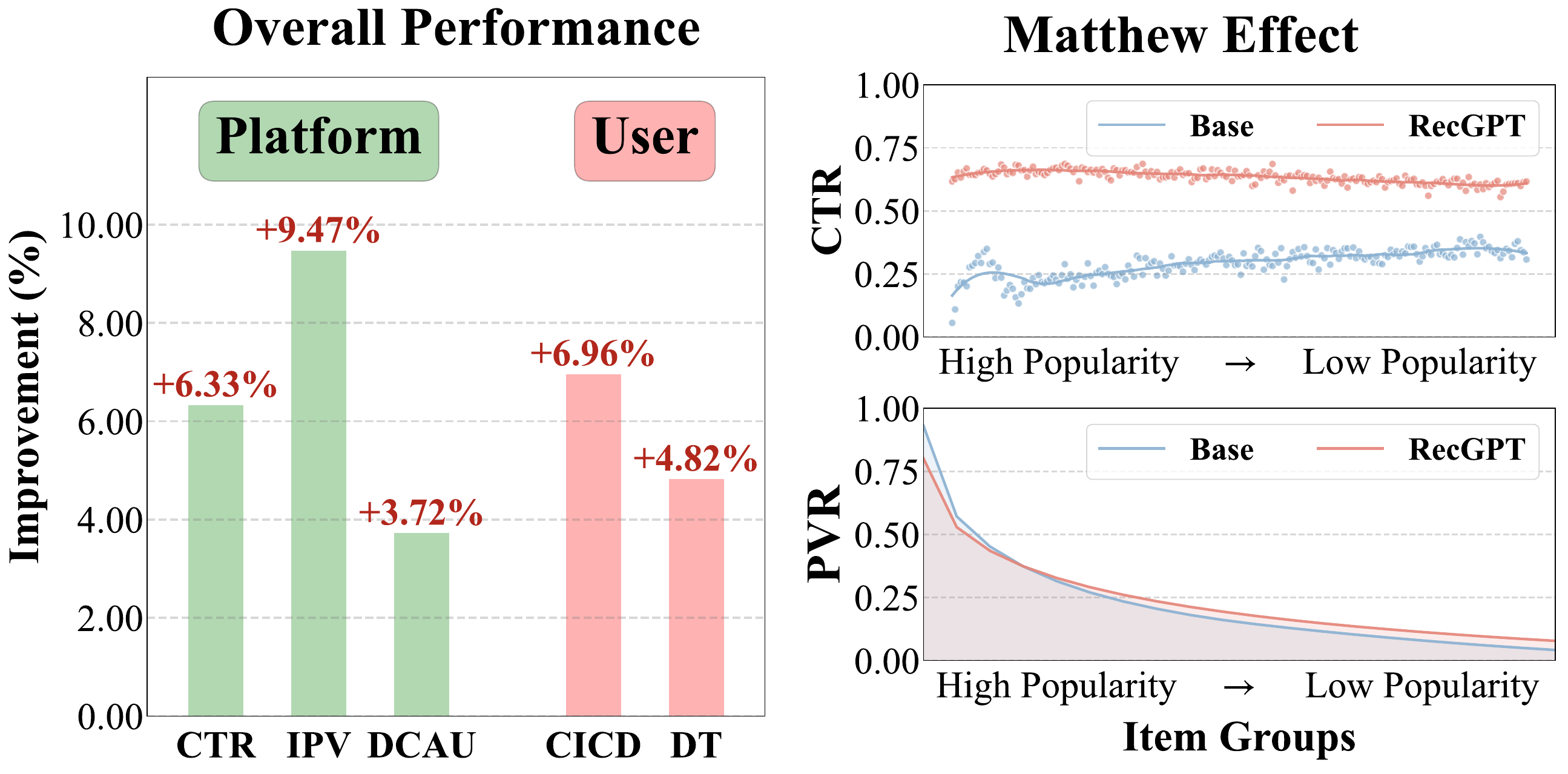}
    \vspace{-1em}
    \caption{Online performance of RecGPT in the ``Guess What You Like'' scenario on Taobao APP's homepage. The left figure shows the overall performance improvements of RecGPT compared to the baseline system across key metrics, including Click Through Rate (CTR), Item Page View (IPV), Daily Click Active Users (DCAU), per-user Clicked Item Category Diversity (CICD), and user Dwell Time (DT). The right figure illustrates the CTR and Page View Rate (PVR) distributions of RecGPT and the baseline system across product groups of different popularity levels. For the purpose of protecting business confidentiality, the commercial metrics (such as CTR and PVR) have been normalized.}
    \label{fig:abs}
\end{figure}

\newpage
\setcounter{tocdepth}{2} 

\tableofcontents 

\newpage
\section{Introduction}
Recommender systems have become pervasive across modern digital ecosystems, ranging from e-commerce portals such as Taobao and Amazon to content platforms like YouTube and TikTok, fundamentally reshaping how people discover and consume information \citep{lu2012recommender}. An ideal recommender system should match a user’s (often implicit) intent with the most relevant item or content, allowing the user to obtain the maximum experience value with minimal effort~\citep{resnick1997recommender}. When this alignment is achieved, it forms a positive feedback loop that benefits all stakeholders: users enjoy satisfying experiences, merchants see increased sales, and platforms gain sustained traffic and revenue. The crux of realizing this vision, however, lies in accurately inferring and modeling the user’s true intent behind the long and rapidly changing behavior traces.

Over the past two decades, both academia and industry have pursued this vision through relentless optimization of \textit{feature engineering} and \textit{model architecture}. Feature representations have progressed from hand-crafted statistics to sequential and cross features, and most recently to ultra-long behavior modeling~\citep{schifferer2020tutorial}. Model architectures have advanced from factorization machines~\citep{rendle2010factorization} to deep matching networks~\citep{zhang2019deep}, graph neural models~\citep{wu2022graph}, and the latest generative Transformer backbones~\citep{deng2025onerec}. Although these efforts have delivered remarkable business gains, they remain fundamentally limited by the co-occurrence patterns found in historical logs—they essentially ``\textit{learn clicks from clicks}''. Lacking an explicit understanding of user interest, such log-fitting methods tend to reinforce what similar users have already consumed, amplifying \textit{Filter Bubble} effects~\citep{nguyen2014exploring,wang2022user} and further marginalizing long-tail sparsity (i.e., \textit{Matthew Effects}~\citep{liu2021examining,gao2023alleviating}). Bridging this gap calls for a new modeling paradigm that can look beyond surface-level correlations and reason about the motivations that drive user behavior.

The recent advent of large language models (LLMs)~\citep{zhao2023survey}, especially those with strong reasoning capabilities, has opened a promising pathway to transcend the limitations of purely log-fitting recommendation. Thanks to their broad world knowledge, fine-grained semantic understanding, and step-wise reasoning abilities, LLMs can help accurately and comprehensively analyze user potential interests and explicitly reason about why a user may want an item.
Although a growing body of work has begun to use LLMs to enhance recommender systems, most studies are limited to small, offline benchmarks, and cannot be applied in real-world recommendation environments~\citep{zhao2024recommender, wu2024survey}.
How to effectively integrate LLMs into large-scale industrial recommender systems to truly understand and mine user intent—so as to overcome the limitations of log-fitting recommendation—remains largely underexplored.

To fill this gap, we introduce \textbf{RecGPT}, a production-scale framework that integrates three reasoning LLMs into the core of an industrial recommendation pipeline, forming a closed loop of ``User Interest Mining $\rightarrow$ Item Tag Prediction $\rightarrow$ Item Retrieval $\rightarrow$ Explanation Generation'' (Figure~\ref{fig:overview}).
Specifically, RecGPT first employs a \textit{User-Interest LLM}  to comprehensively analyze users' lifelong behavior history and explicitly generate a concise, natural-language profile of current interests. A second \textit{Item-Tag LLM} then reasons over those interests to generate fine-grained item tags that describe the items the user is most likely seeking. These tags are injected into the item-retrieval stage, expanding the conventional user–item dual-tower matcher into a user–item–tag tri-tower architecture. Consequently, only items that align with the inferred user intent are passed on to the downstream ranking and re-ranking cascade. By turning user behavior logs into a dynamically updated intent signature, RecGPT reshapes candidate generation from collaborative filtering to interest-enhanced process, improving recall relevance and long-tail coverage without changing the downstream infrastructure. Finally, a \textit{Recommendation-Explanation LLM} attaches cached, natural-language user-friendly explanations to final recommended items, completing the loop from intent discovery to transparent delivery.

Our main contributions are summarized as follows:

\faStar\; 
RecGPT has been fully deployed online in the ``\textit{Guess What You Like}'' scenario on Taobao APP's homepage, achieving significant performance improvements across multiple stakeholders. From the user perspective, our system enhances CICD by \textcolor{red}{6.96\%} and DT by \textcolor{red}{4.82\%}, indicating effective discovery of users' latent and diverse interests beyond historical interaction patterns, effectively breaking through information bubbles and expanding recommendation boundaries. For merchants and the platform, notable improvements are observed in CTR (\textcolor{red}{+6.33\%}), IPV (\textcolor{red}{+9.47\%}), and DCAU (\textcolor{red}{+3.72\%}), reflecting substantial commercial value. Additionally, RecGPT effectively alleviates the Matthew effect by providing more equitable exposure opportunities across diverse merchants, ultimately establishing a \textbf{win-win-win ecosystem} for users, merchants, and the platform.

\faStar\; 
Unlike traditional collaborative filtering approaches that ``learn clicks from clicks'', we leverage LLMs' world knowledge and reasoning capabilities to explicitly mine users' latent interests from behavioral patterns, shifting from surface-level correlations to deep profile analysis and preference modeling. To the best of our knowledge, \textit{\textbf{we are the first to deploy a reasoning-enhanced hundred-billion-scale recommendation foundation model in industrial applications serving over a billion consumers and items}}, which powerfully validates and advances the practical potential and value of large language models for recommender systems.

\faStar\; 
To enable effective LLM integration in large-scale industrial recommender systems, we develop a systematic multi-stage training framework that addresses the unique challenges of adapting general-purpose LLMs to recommendation-specific tasks. Our approach progresses from reasoning-enhanced pre-alignment to self-training evolution, leveraging LLM-as-a-Judge capabilities for automated data quality curation and model evaluation. This framework enables \textbf{a progressive transition from manual expert review to a Human-LLM cooperative judge system}, significantly accelerating model iteration cycles while maintaining rigorous quality standards.





\section{RecGPT Workflow}
\label{sec:RecGPT_Workflow}
\begin{figure}[t]
    \centering
    \includegraphics[width=\textwidth]{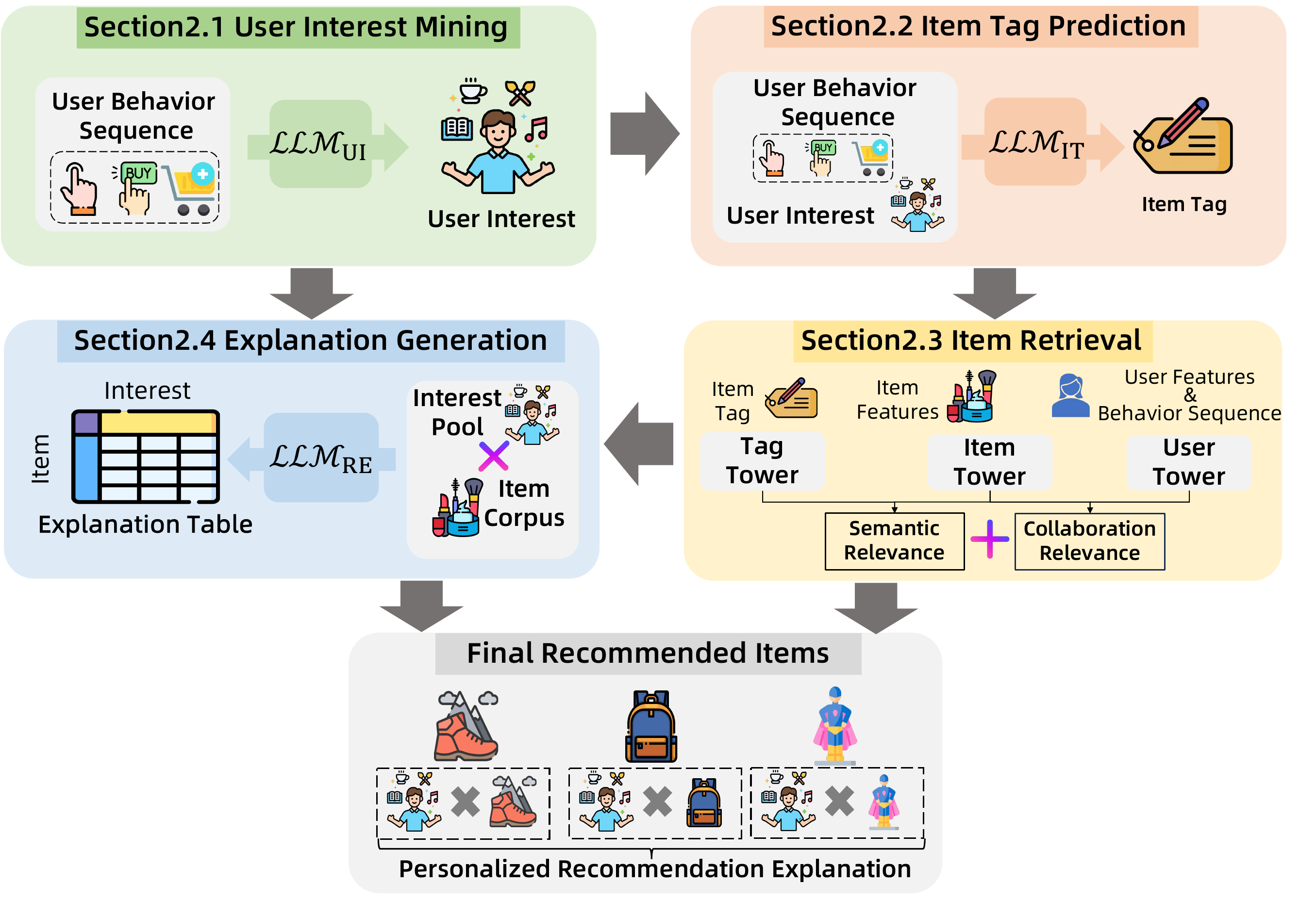}
    \caption{The overall workflow of RecGPT. $\mathcal{LLM}_{\text{UI}}$, $\mathcal{LLM}_{\text{IT}}$, and $\mathcal{LLM}_{\text{RE}}$ represent LLMs for user interest mining, item tag prediction, and recommendation explanation generation, respectively. RecGPT employs ``user interest mining $\to$ item tag prediction'' to identify potential items matching user preferences, then performs item retrieval through tag-aware semantic relevance combined with behavioral collaboration to map inferred tags to specific in-domain items. Finally, $\mathcal{LLM}_{\text{RE}}$ generates user-friendly recommendation explanations based on the retrieval results and user interests.}
    \label{fig:overview}
\end{figure}

In this section, we present the overall workflow of RecGPT, as illustrated in Figure~\ref{fig:overview}. The core idea of RecGPT is to leverage large language models to empower different stages of the recommendation pipeline, including user interest understanding, item prediction, and generating user-friendly recommendation explanations for final results. We introduce three corresponding LLM modules: $\mathcal{LLM}_{\text{UI}}$ serves user interest mining tasks, $\mathcal{LLM}_{\text{IT}}$ handles item tag prediction tasks, and $\mathcal{LLM}_{\text{RE}}$ generates recommendation explanations.
Furthermore, to map the items (referred to as \textit{item tags} in this paper) predicted by $\mathcal{LLM}_{\text{IT}}$ to specific items within the in-domain item corpus, we propose a tag-aware semantic relevance retrieval method. 
This approach leverages the deep semantic understanding derived from LLM-generated item tags, which captures user intent through reasoning-based analysis rather than surface-level feature matching. By integrating these LLM-driven semantic insights with traditional collaborative filtering signals, our item retrieval method effectively balances semantic relevance and collaborative patterns, enabling both exploration of users' potential diverse preferences and exploitation of their established behavioral patterns.

Overall, the RecGPT workflow consists of the following components:
\begin{itemize}[topsep=0em, itemsep=0.5em, label=$\blacklozenge$]
\item \textbf{User Interest Mining} (Section~\ref{sec:User_Interest_Mining}):
Through $\mathcal{LLM}_{\text{UI}}$, we conduct explicit interest mining on users' lifelong multi-behavior sequences to identify diverse user interest patterns.
\item \textbf{Item Tag Prediction} (Section~\ref{sec:Item_Tag_Prediction}):
Based on user interest mining results, we use $\mathcal{LLM}_{\text{IT}}$ to predict item tags that represent users' potential preference distributions.
\item \textbf{Item Retrieval} (Section~\ref{sec:Item_Retrieval}):
The tag-aware semantic retrieval method maps predicted tags to specific items while incorporating user behavioral collaborative signals to balance semantic and collaborative relevance.
\item \textbf{Recommendation Explanation Generation} (Section~\ref{sec:Explanation_Generation}):
$\mathcal{LLM}_{\text{RE}}$ synthesizes user interests and recommended items to generate personalized and user-friendly explanations that resonate with individual user preferences, improving system transparency and user experience.
\end{itemize}

In contrast to traditional recommendation algorithms that rely on latent features and final user feedback for optimization, RecGPT employs explicit text-based modeling through large language models across pipeline stages. This approach provides two key advantages: First, it enables interpretable monitoring of intermediate processes and model performance at each stage. Second, it facilitates expert knowledge integration through process-level supervision, allowing targeted optimization of individual components. By decomposing the workflow into manageable sub-tasks with clear input-output relationships, this methodology simplifies end-to-end optimization while enabling independent evaluation and refinement.

\subsection{User Interest Mining}
\label{sec:User_Interest_Mining}

The fundamental goal of recommender systems is to understand users' personalized preferences through their historical interaction behaviors and achieve item product recommendations. However, existing recommendation algorithms rely solely on fixed, statistical implicit user features, making it difficult to explicitly model users' dynamic and complex interests. To overcome these fundamental limitations, we introduce \textbf{Generative User Profiling}, a novel approach that harnesses the powerful reasoning capability of LLMs to revolutionize user interest modeling.
However, despite their promising potential in natural language understanding, several key challenges hinder LLMs' effectiveness in user interest mining:

\textbf{(1) Context Window Limitations.} Real-world user behavioral histories in recommender systems exhibit vast scale and complexity. Within Taobao's e-commerce platform, users possess over $37$k historical behavior records on average. These extensive behavioral sequences pose significant challenges to current LLMs limited by $128$k-token context windows.

\textbf{(2) Domain Knowledge Gaps.} Despite broad world knowledge, LLMs lack specialized understanding of domain-specific features in platforms like Taobao. This knowledge gap hinders the models' ability to effectively extract and abstract user interests from raw interaction data at an expert level.

To overcome these challenges, we first develop \textbf{\textit{Reliable Behavioral Sequence Compression}} to preserve critical temporal information while reducing input length, enabling better adaptation to LLM context window constraints (Section~\ref{sec:User_Interest_Compression}). Building upon this, we present a multi-stage \textbf{\textit{Task Alignment}} framework for the \textit{User-Interest LLM} $\mathcal{LLM}_{\text{UI}}$ to enhance domain-specific user interest mining capabilities (Section~\ref{sec:User_Interest_LLM}).

\subsubsection{Reliable Behavioral Sequence Compression}
\label{sec:User_Interest_Compression}
To enhance the reliability and effectiveness of user behavior sequences, we first employ a \textit{Reliable Behavior Extraction} method to filter out noise and redundant information from user behaviors. Furthermore, to accommodate ultra-long user behavior sequences within the context window limitations of LLMs, we develop a \textit{Hierarchical Behavior Compression} method for heterogeneous behavior compression.

\paragraph{Reliable Behavior Extraction.} To ensure that user interaction behaviors accurately reflect genuine user interests, we first extract reliable signals from large-scale, multi-source, multi-behavior user sequences as the data foundation for user interest modeling. We define the following reliable behaviors (using Taobao e-commerce platform as an example):

\begin{itemize}[topsep=0em, itemsep=0.5em, label=$\blacklozenge$]
\item \textbf{Intentional Feedback Behaviors} include high-engagement actions such as ``favorites'', ``purchases'', ``add-to-cart'', and deliberate click behaviors like ``detailed product views'' and ``reviews reading'', demonstrating strong user interest through direct engagement actions or focused clicking patterns that reflect deep attention to item details and provide robust signals for modeling user preferences and purchase intentions.
\item \textbf{Search Behaviors} comprise actions such as ``search queries'' for product discovery. These behaviors represent deliberate exploration efforts, revealing user intentions toward specific product categories or attributes.
\end{itemize}
Note that we exclude ordinary product clicking behaviors, as these actions may contain considerable noise and are less effective at reflecting user interests compared to the defined reliable behaviors.

\paragraph{Hierarchical Behavior Compression.}
To accommodate the ultra-long user behavior sequences, we develop a hierarchical compression method that compresses multi-source heterogeneous behaviors into a unified sequence format. 
Specifically, we employ compression strategies at both the item level and sequence level, which are designed to reduce the input length while preserving essential information for interest mining.

\textbf{(1) \textit{Item-level Compression.}} Considering that raw item information contains substantial redundant and irrelevant details, using this raw data as input to LLMs would result in low information density and excessive token consumption. Therefore, we first compress the relevant information for each item. Here, we prompt the LLM to compress detailed item information while preserving core attributes such as item name, category, brand, and other essential features.

\textbf{(2) \textit{Sequence-level Compression.}} After compressing individual item information, we further compress the user's behavioral sequences through a two-step aggregation process. We first partition the user's behavior sequence into different time periods (daily partitions for behaviors within one month, monthly partitions for behaviors spanning multiple months, and yearly partitions for behaviors exceeding one year).
Our compression method operates through two complementary aggregation steps:
(1) \textbf{Step 1: Temporal-Behavioral Aggregation.} We use ``time-behavior type'' pairs as keys to group all items that the user interacted with during specific time periods through specific behaviors.
(2) \textbf{Step 2: Item-based Reverse Aggregation.} We then reverse this mapping by using item sequences as keys to aggregate their corresponding time-behavior type combinations. For example, items that frequently appear together across different time periods and behaviors are grouped, with their associated temporal-behavioral contexts preserved.
This dual aggregation process produces compressed behavioral sequences in the following format (specific cases can be referred to in Section~\ref{sec:Case_Studies}):

``\textcolor{ForestGreen}{$\text{Time}_1$} \textcolor{Purple}{($\text{Behavior}_1$, $\text{Behavior}_2$, $\ldots$)}, 
\textcolor{ForestGreen}{$\text{Time}_2$} \textcolor{Purple}{($\text{Behavior}_1$, $\text{Behavior}_3$, $\ldots$)}, $\ldots$ $|$ 
\textcolor{YellowOrange}{$\text{Item}_1$, $\text{Item}_2$, $\ldots$}''

\noindent
This representation efficiently captures both temporal behavioral patterns and item co-occurrence relationships while significantly reducing overall prompt sequence length.

Through the above behavior extraction and compression process, we obtain multi-source heterogeneous user behavior sequences $\mathcal{B}_u = [B_{u,1}, B_{u,2}, \ldots, B_{u,|\mathcal{B}_u|}]$ with higher information density that are both refined and reliable, where each behavior $B_{u,i}$ contains multiple interactions within the same time period, along with their corresponding interaction behavior types and associated item information. 
We empirically demonstrate that the proposed reliable behavior compression method effectively accommodates $\mathbf{98\%}$ of user behaviors within the $128$k-token context window of large language models, compared to only 88\% coverage achieved by uncompressed sequences. Furthermore, this compression approach improves interest inference efficiency by $\mathbf{29\%}$, significantly reducing both inference time and computational costs while maintaining complete behavior representation.

\begin{figure}[t]
    \centering
    \includegraphics[width=\textwidth]{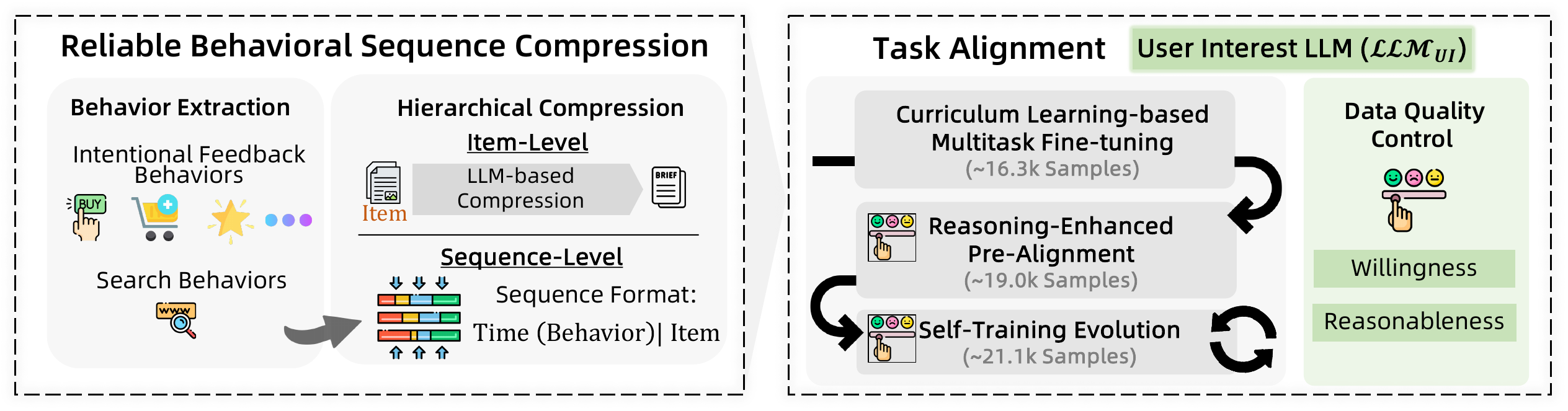}
    \caption{Illustration of the user interest mining module. The left figure demonstrates the compression processing of lifelong user behavioral sequences, including behavior extraction and hierarchical behavior compression. The right figure shows the multi-stage task alignment framework for user interest mining and data quality control standards.}
    \label{fig:interest}
\end{figure}

\subsubsection{Task Alignment for User Interest Mining}
\label{sec:User_Interest_LLM}

To enhance User Interest LLM $\mathcal{LLM}_{\text{UI}}$ capability in interest mining task, we design a multi-stage task alignment framework with the following stages to develop a human-aligned $\mathcal{LLM}_{\text{UI}}$:

\textbf{Stage 1: Curriculum Learning-based Multi-task Fine-tuning.}
We first design $16$ preparatory subtasks (containing $16.3$k training samples) to enhance general-purpose LLMs' domain-specific foundational abilities. These subtasks develop key competencies across multiple dimensions, such as key information extraction, complex user profile analysis, and causal reasoning.
To ensure stable capability enhancement, inspired by curriculum learning principles~\citep{bengio2009curriculum,wang2021survey,soviany2022curriculum,pentina2015curriculum}, we organize subtasks through topological sorting based on difficulty levels and dependency relationships, progressively guiding the model to master complex tasks. 
Relevant experimental details are provided in Appendix~\ref{sec:Curriculum_Learning_Multi_task_Fine_tuning}.

\textbf{Stage 2: Reasoning-Enhanced Pre-alignment.}
We leverage the advanced reasoning capabilities of DeepSeek-R1~\citep{guo2025deepseek} to generate high-quality training data for interest mining. Through careful manual curation, we distill the initial $90.0$k generated samples into a refined $19.0$k high-quality dataset. This dataset serves as the foundation of knowledge distillation, allowing the user interest LLM $\mathcal{LLM}_{\text{UI}}$ to achieve performance comparable to the teacher model via pre-alignment fine-tuning.

\textbf{Stage 3: Self-Training Evolution.}
To further enhance the model's capability ceiling, we propose a self-training paradigm that enables continuous self-evolution. In this stage, the model generates its own training data and uses these self-generated samples for iterative optimization, creating a feedback loop for capability improvement. During this self-training process, we collect $21.1$k high-quality samples that drive the model's evolution. To efficiently filter these self-generated outputs and evaluate model performance at low cost, we adopt a \textit{Human-LLM collaborative paradigm} with \textit{LLM-as-a-Judge} capabilities for data quality control and assessment. This collaborative framework significantly improves curation efficiency while reducing manual annotation costs. We provide a comprehensive introduction to the Human-LLM collaborative judge system in Section~\ref{sec:Human_LLM_Cooperation}.

In what follows, we focus on explaining the \textbf{Prompt Engineering} strategy and \textbf{Data Quality Control} protocols for the interest mining task. Additionally, we present extensive \textbf{Human Evaluation Experiments} to demonstrate the effectiveness of our self-training method and provide practical \textbf{Online Deployment} details.

\paragraph{Prompt Engineering.}
Our carefully designed prompt template takes compressed behavior sequences (\textit{cf.} Section~\ref{sec:User_Interest_Compression}) and personal attributes as user information, instructing the LLM to generate diverse yet precise interest profiles. To improve generation accuracy, the template incorporates \textit{Chain-of-Thought} (CoT) reasoning that guides the model through explicit logical steps instead of direct interest prediction. The prompt template structure is shown in Prompt~\ref{box:interest-mining}, where the placeholders \textcolor{BlueViolet}{\textbf{\{User Attributes\}}} and \textcolor{BlueViolet}{\textbf{\{Compressed Reliable Behavioral Sequences\}}} are instantiated with user-specific contextual data. 
Detailed specifications for \textcolor{Maroon}{\textbf{\{Other Interest Mining Requirements\}}} and \textcolor{Maroon}{\textbf{\{Other Constraints\}}} are provided in Appendix~\ref{box:app-interest-mining} due to space limitations. The \textcolor{Maroon}{\textbf{\{Matched Interest Pool\}}} is filled with a dynamic collection of matched interests.

\begin{tcolorbox}[promptbox=interest_colframe/interest_colback, title={User Interest Mining Prompt Template},label=box:interest-mining,breakable]
\colorbox{green!20}{\textbf{\# Role}}\\
You are a shopping guide for an e-commerce platform. Based on users' behavioral history, you need to accurately and comprehensively analyze their potential interests and preferences.\\
\colorbox{blue!20}{\textbf{\# Input}}\\
\textbf{User Attribute}: \textcolor{BlueViolet}{\textbf{\{User Attributes\}}}

\textbf{User Behavioral Information}:
\textcolor{BlueViolet}{\textbf{\{Compressed Reliable Behavioral Sequences\}}}\\
\colorbox{red!20}{\textbf{\# Mandatory Requirements}}\\
\textbf{Task Requirements}
\textcolor{Maroon}{\textbf{\{Interest Mining Requirements\}}}\\
\textbf{Task Constraints}: \textcolor{Maroon}{\textbf{\{Constraints\}}}\\
\colorbox{orange!20}{\textbf{\# Preset Interest List}}\\
\textcolor{Maroon}{\textbf{\{Matched Interest Pool\}}}\\
\colorbox{yellow!20}{\textbf{\# Output Format}}\\
\textit{(Detailed output format requirements)}
\end{tcolorbox}

Following the prompt template design, we formalize the user interest mining process as follows. Given a target user's attributes $\mathcal{A}_u$ extracted from user-provided information (\textit{e.g}, age, gender, location), compressed behavioral sequence $\mathcal{B}_u$, and the pre-configured matched candidate interest pool $\mathcal{I}_m$, we leverage the User-Interest LLM $\mathcal{LLM}_{\text{UI}}$ to perform inference:
\begin{equation}
\mathcal{I}_u = \mathcal{LLM}_{\text{UI}}(\mathcal{A}_u, \mathcal{B}_u, \mathcal{I}_m \mid \mathcal{P}_{\text{UI}}),
\end{equation}
where $\mathcal{P}_{\text{UI}}$ represents the user interest mining prompt template, and $\mathcal{I}_u = \{I_{u,1}, I_{u,2}, \ldots, I_{u,|\mathcal{I}_u|}\}$ denotes the set of potential user interests inferred from the given user information. This approach enables the model to synthesize user attributes, behavioral patterns, and candidate interests through CoT-based reasoning to mine personalized interest profiles.

\paragraph{Data Quality Control.}
To mitigate potential LLM hallucination and inaccuracies, we employ multi-dimension reject sampling for training data preparation. The evaluation criteria for correct (\correct) and incorrect (\wrong) interests include 2 dimensions:

\begin{table}[t]
    \centering
    \caption{Criteria for Evaluating Model-Generated Interests. \textbf{Correct} indicates interests that meet both Willingness and Strong Correlation criteria, while \textbf{Incorrect} includes four categories: Necessity, Weak Correlation, No Correlation and Hallucination. ``Both \correct $\rightarrow$ \correct'' means when both criteria are satisfied, the interest is considered correct, while ``Any \wrong $\rightarrow$ \wrong'' means if any criterion is violated, the interest is deemed incorrect.}
    \label{tab:reject_interest_criteria}
    \renewcommand{\arraystretch}{1.2} 
    \begin{tabularx}{\textwidth}{ccXX}
        \toprule
        \multicolumn{1}{c}{\textbf{Label}} & \multicolumn{1}{c}{\textbf{Evaluation Criteria }} & \multicolumn{1}{c}{\textbf{Example}} & \multicolumn{1}{c}{\textbf{Why \correct or \wrong}} \\
        \midrule
        \multirow{2}{*}[-3ex]{\makecell{\textbf{Correct}\\(Both \correct $\rightarrow$ \correct)}} 
            & \multirow{1}{*}[-1.5ex]{\makecell{Spontaneity \\ (Willingness)}} &  \multirow{1}{*}[-2ex]{[Interest] \textit{tennis.}} & Tennis is an interest from affection. (Willingness\ \correct) \\
            & \multirow{1}{*}[-1.5ex]{\makecell{Strong Correlation \\ (Reasonableness)}} & [Reason] \textit{User purchased a tennis racket.} & Reason and the interest are related reasonably. (Strong Correlation\ \correct)  \\
        \midrule
        \multirow{8}{*}[-8ex]{\makecell{\textbf{Incorrect}\\(Any \wrong $\rightarrow$ \wrong)}} 
            & \multirow{1}{*}[-3ex]{\makecell{Necessity \\ (Willingness)}}   
            & [Interest] \textit{Purchasing household necessities.}& \multirow{2}{=}[-3ex]{Purchasing household items is a daily need, not a hobby.} \\
            && [Reason] \textit{Most purchases are daily cleaning items.} & \\
            \cline{2-4}
            & \multirow{1}{*}[-1.5ex]{\makecell{Weak Correlation \\ (Reasonableness)}} 
            &  [Interest] \textit{Yoga.}& \multirow{2}{=}[-1ex]{May indicate clothing hobby, not yoga preference.}  \\
            && [Reason] \textit{Purchases include yoga pants.} &\\
            \cdashline{2-4}
              & \multirow{1}{*}[-1.5ex]{\makecell{No correlation \\ (Reasonableness)}} & [Interest] \textit{Model making.} &  \multirow{2}{=}[-1ex]{Interest and reasoning are unrelated.}\\
             & & [Reason] \textit{User searched for many blankets.} & \\
            \cdashline{2-4}
            & \multirow{1}{*}[-1.5ex]{\makecell{Hallucination \\ (Reasonableness)}} 
            & [Interest] \textit{Smart home. } & \multirow{2}{=}[-1ex]{User history has no smart home-related activities.} \\
            & & [Reason] \textit{User purchased smart home accessories.} & \\
        \bottomrule
    \end{tabularx}
\end{table}

\begin{itemize}[itemsep=0.5em, topsep=0.5em, label=$\blacklozenge$]
    \item \textbf{Willingness.} This criterion evaluates whether the identified interests genuinely reflect voluntary user preferences rather than external obligations. Genuine interests embody intrinsic motivation and active exploration, distinguishing them from necessity-driven behaviors.
    \begin{itemize}[itemsep=0.5em, topsep=0.5em]
        \item \correct\; \textbf{Spontaneity.} The interest originates from voluntary affection and personal choice, which is driven by curiosity, passion, or fulfillment that aligns with one's values, experiences, or aspirations. Such interests reflect authentic user preferences and intrinsic motivation.
        \item \wrong\; \textbf{Necessity.} Behaviors misclassified as interests may actually stem from external pressures, survival needs, or situational requirements (e.g., career-related skill acquisition). These represent functional demands rather than genuine personal interests.
    \end{itemize}
    \item \textbf{Reasonableness.} This criterion ensures that inferred user interests have sufficient behavioral evidence for support, maintaining reliability and accuracy in interest identification. We establish four correlation degrees to evaluate interest reasonableness:
    \begin{itemize}[itemsep=0.5em, topsep=0.5em]
    \item \correct\; \textbf{Strong Correlation.} The inferred interest and observed behaviors demonstrate clear, logical connections with compelling evidence. The behavioral patterns strongly support the interest inference with minimal ambiguity.
    \item \wrong\; \textbf{Weak Correlation.} User behaviors show partial connection to the inferred interest but lack sufficient evidence. For example, purchasing a tennis skirt may provide some indication of interest in tennis but cannot conclusively establish this preference.
    \item \wrong\; \textbf{No Correlation.} User behaviors exhibit no meaningful connection to the inferred interest. For instance, purchasing ``The Prince of Tennis'' merchandise does not indicate interest in tennis sports, as it likely reflects interest in the anime rather than the sport.
    \item \wrong\; \textbf{Hallucination.} The generated interest has no behavioral evidence, representing unfounded associations fabricated by the LLM.
    \end{itemize}
\end{itemize}

Using the above data quality control protocol, we retain data exhibiting strong user intent and clear correlations for training, while filtering out data with weak intent, poor correlations, or hallucinations as low-quality samples that could introduce noise and bias into the learning process. Through this rigorous data curation process, we ensure consistently high data quality across both pre-alignment and self-training stages, thereby improving model performance on user interest mining tasks.

\paragraph{Human Evaluation Experiments.}
\begin{table}[h]
\centering
\caption{Human-evaluated pass rates for different models on user interest mining task. The best performance is highlighted in \textbf{bold}.}
\label{tab:human_evaluation_interest_mining}
\begin{tabular}{ccccc}
\toprule
\textbf{Model} & \textbf{DeepSeek-R1} & \textbf{Qwen3-Base} & \textbf{Qwen3-SFT} & \textbf{TBStars-SFT} \\
\midrule
\textbf{Pass Rate (\%)} & 70.00 & 59.74 & $\mathbf{77.28}$ & 74.39 \\
\bottomrule
\end{tabular}
\end{table}

To validate the effectiveness of our multi-stage alignment framework, we conduct human evaluation on user interest mining performance across different models. We evaluate two foundation LLMs: DeepSeek-R1 and Qwen3-14B (hereafter referred to as \textbf{Qwen3-Base}), alongside our multi-stage aligned models: Qwen3 (hereafter referred to as \textbf{Qwen3-SFT}) and TBStars-42B-A3.5 (hereafter referred to as \textbf{TBStars-SFT}), a sparse Mixture-of-Experts (MoE) large language model internally developed by Taobao that activates only 3.5B parameters per inference. The generated interest is considered ``passed'' only when it satisfies all data quality evaluation criteria outlined before, including both willingness and reasonableness dimensions.

As shown in Table~\ref{tab:human_evaluation_interest_mining}, DeepSeek-R1 achieves a 70.00\% pass rate, significantly outperforming Qwen3-Base at 59.74\%. The performance gap demonstrates that reasoning-enhanced LLMs possess superior contextual understanding capabilities for mining user interests from ultra-long behavioral sequences.
Furthermore, our multi-stage aligned Qwen3-SFT achieves the highest performance at 77.28\%, substantially outperforming its base counterpart. The experimental results validate the effectiveness of our alignment framework in enhancing domain-specific interest mining capabilities.

For practical online deployment, TBStars-SFT achieves 74.39\% pass rate after full-parameter fine-tuning with the collected high-quality data, demonstrating significant improvement over both baseline models (\textit{i.e.}, DeepSeek-R1 and Qwen3-Base) while maintaining superior performance-efficiency balance due to its sparse architecture. This characteristic makes it particularly suitable for large-scale online recommendation scenarios where both accuracy and inference speed are critical.

\paragraph{Online Deployment.} We utilize the model $\mathcal{LLM}_{\text{UI}}$ offline to predict users' interest preferences, with an average of $\mathbf{16.1}$ predicted interests per user. During online deployment, we perform iterative model optimization and refresh user interests every two weeks to ensure the timeliness of the user interest, while precisely capturing the dynamic changes in users' personalized interests.


\subsection{Item Tag Prediction}
\label{sec:Item_Tag_Prediction}
In this section, we explore how to leverage large language models to guide item tag prediction based on inferred user profiles. To adapt the world knowledge of LLMs to the specific product domain, similar to the User Interest LLM $\mathcal{LLM}_{\text{IT}}$, we first perform multi-stage \textbf{Task Alignment} to ensure $\mathcal{LLM}_{\text{IT}}$ can effectively understand and process product-related contextual information (Section~\ref{sec:Item_Tag_Prediction_Task_Alignment}). Next, we introduce a \textbf{Incremental Learning} method that enables the model to continuously adapt to evolving user interests and new product trends (Section~\ref{sec:Item_Tag_Prediction_Incremental_Learning}).

\subsubsection{Task Alignment for Item Tag Prediction}
\label{sec:Item_Tag_Prediction_Task_Alignment}
\begin{figure}[t]
    \centering
    \includegraphics[width=\textwidth]{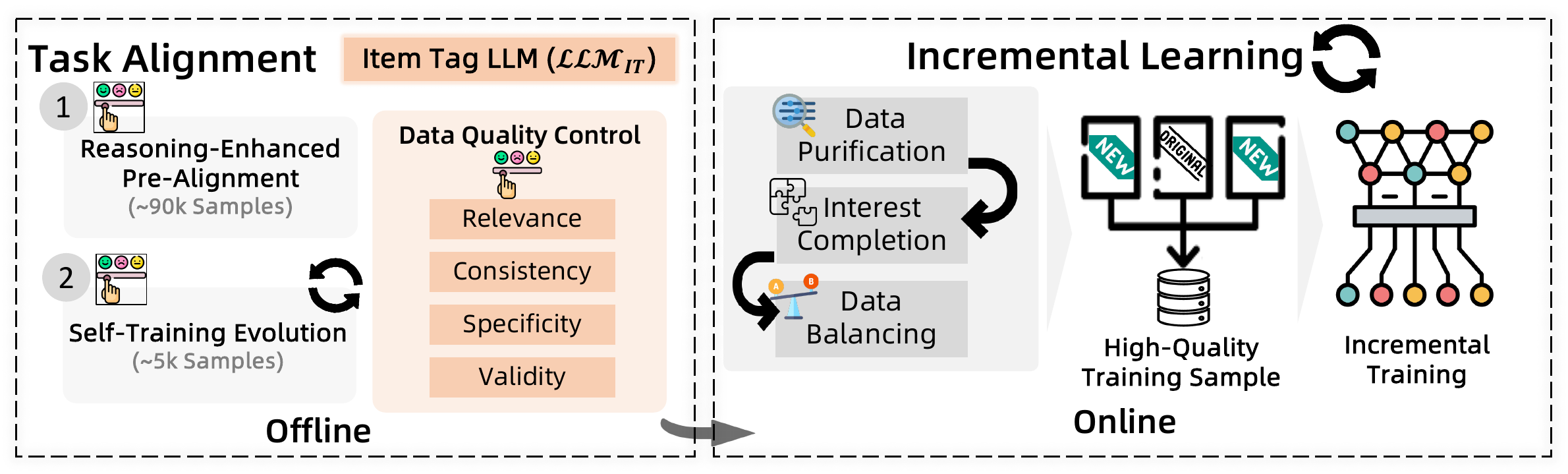}
    \caption{Illustration of the item tag prediction module. The left figure demonstrates the two-stage alignment process for the item tag LLM and data quality control standards, while the right figure shows data cleaning and incremental learning based on real online user feedback data.}
    \label{fig:tag}
\end{figure}

While foundation large language models show impressive general capabilities, they prove inadequate when directly applied to personalized item prediction tasks due to the domain-specific requirements of recommender systems. To overcome this limitation, we adopt a two-stage alignment process similar to the Item-Tag LLM approach. This process employs \textbf{Reasoning-Enhanced Pre-Alignment} and \textbf{Self-Training Evolution} to enhance $\mathcal{LLM}_{\text{IT}}$ with domain-aware product understanding. Here, we focus on the \textbf{Prompt Engineering} and \textbf{Data Quality Control} protocols specifically designed for the item tag prediction task.
Furthermore, we provide extended \textbf{Human Evaluation Experiments} to demonstrate the effectiveness of our alignment approach in improving model performance.

\paragraph{Prompt Engineering.}
We require the Item-Tag LLM $\mathcal{LLM}_{\text{IT}}$ to predict tag sets in the ``\textbf{\textcolor{Salmon}{Modifier} + \textcolor{RubineRed}{Core-Word}}'' format (\textit{e.g}, ``\textcolor{Salmon}{Outdoor waterproof non-slip} \textcolor{RubineRed}{hiking boots}''), based on provided user profile information and multi-behavior interaction sequences (such as clicks, purchases, searches). We employ CoT-based tag reasoning to fully leverage the reasoning capabilities of large language models.
Additionally, we incorporate the following constraints in our prompts to ensure the generated tag sets meet the practical requirements of recommender systems as follows:

$\blacklozenge$\;
\textbf{Interest Consistency}: Generated tags are constrained to maintain alignment with user interests, thereby preventing recommendations that contradict established user preference profiles.

$\blacklozenge$\;
\textbf{Diversity Enhancement}: A minimum of 50 tags is enforced to guarantee diverse recommendation across broad categories, mitigating filter bubble phenomena.

$\blacklozenge$\;
\textbf{Semantic Precision}: Tag generation is restricted to semantically focused descriptions, eliminating vague or overly broad categorizations that compromise recommendation accuracy and user experience quality.

$\blacklozenge$\;
\textbf{Temporal Freshness}: The generated tag should prioritize novel product categories while systematically avoiding repetitive recommendations of recently engaged items, ensuring both temporal relevance and diversified suggestions.

$\blacklozenge$\;
\textbf{Seasonal Relevance}: Temporal context is integrated into the tag generation process to produce seasonally appropriate recommendations aligned with the provided timestamp, enhancing user satisfaction through contextually aware suggestions.

Through the above multi-constraint prompting strategies, $\mathcal{LLM}_{\text{IT}}$ generates a list of triplets containing \textbf{(Tag, Associated Interest Preference, Rationale)} for subsequent item retrieval. The detailed tag prediction prompt template structure is presented in Prompt~\ref{box:tag-prediction}, where \textcolor{BlueViolet}{\textbf{\{User Attributes\}}}, \textcolor{BlueViolet}{\textbf{\{User Interests\}}}, \textcolor{BlueViolet}{\textbf{\{Click Behavior Sequence\}}}, \textcolor{BlueViolet}{\textbf{\{Purchase Behavior Sequence\}}}, \textcolor{BlueViolet}{\textbf{\{Search Behavior Sequence\}}}, and \textcolor{BlueViolet}{\textbf{\{Extra Information\}}} are placeholders for the related user profile and behavior records, which are dynamically filled in during the model inference process.
Regarding \textcolor{Maroon}{\textbf{\{Recommendation Principles\}}}, \textcolor{Maroon}{\textbf{\{Recommendation Requirements\}}}, \textcolor{Maroon}{\textbf{\{Quantity Requirements\}}}, and \textcolor{Maroon}{\textbf{\{Strict Prohibitions\}}} in the prompt, due to space constraints, we omit the detailed descriptions here and provide specifications in the Appendix~\ref{box:app-tag-prediction}.

\begin{tcolorbox}[promptbox=tag_colframe/tag_colback, title={Item Tag Prediction Prompt Template},label=box:tag-prediction,breakable]
\colorbox{green!20}{\textbf{\# Role}}\\
You are a professional product recommendation specialist for the Taobao app.\\
\colorbox{blue!20}{\textbf{\# Input}}\\
\textbf{User Attributes}:
\textcolor{BlueViolet}{\textbf{\{Generated User Attributes\}}}\\
\textbf{User Interests}:
\textcolor{BlueViolet}{\textbf{\{Generated User Interests\}}}\\
\textbf{User Behavior Information}\\
Click Behavior Sequence: \textcolor{BlueViolet}{\textbf{\{Click Behavior Sequence\}}} | 
Purchase Behavior Sequence: \textcolor{BlueViolet}{\textbf{\{Purchase Behavior Sequence\}}} |
Search Behavior Sequence: \textcolor{BlueViolet}{\textbf{\{Search Behavior Sequence\}}} |
Extra Information: \textcolor{BlueViolet}{\textbf{\{Extra Information\}}}\\

\colorbox{red!20}{\textbf{\# Mandatory Requirements}}\\
\textbf{Task Requirements}: 
\textbf{**Recommendation Principles (\correct)**} 
|
\textcolor{Maroon}{\textbf{\{Recommendation Principles\}}}
|
\textbf{**Recommendation Requirements (\correct)**}
|
\textcolor{Maroon}{\textbf{\{Recommendation Requirements\}}}
|
\textbf{**Quantity Requirements (\correct)**}
|
\textcolor{Maroon}{\textbf{\{Quantity Requirements\}}}
|
\textbf{**Strict Prohibitions (\wrong)**}
|
\textcolor{Maroon}{\textbf{\{Strict Prohibitions\}}}\\
\colorbox{yellow!20}{\textbf{\# Output Format}}\\
\textit{(Detailed output format requirements)}
\end{tcolorbox}

Following the prompt template design, we formalize the item tag prediction process as follows. Given the inferred user profile information, including user attributes $\mathcal{A}_u$ and interests $\mathcal{I}_u$, along with user multi-behavior interaction sequences $\mathcal{S}_u$ (comprising click behaviors, purchases, and search queries), we leverage the Item Tag LLM to predict item tags that the user might interact with next:
\begin{equation}\label{eq:it}
\mathcal{T}_u = \mathcal{LLM}_{\text{IT}}(\mathcal{A}_u, \mathcal{I}_u, \mathcal{S}_u \mid \mathcal{P}_{IT}),
\end{equation}
where $\mathcal{P}_{\text{IT}}$ represents the item tag prediction prompt template, and $\mathcal{T}_u = \{T_{u,1}, T_{u,2}, \ldots, T_{u,|\mathcal{T}_u|}\}$ denotes the predicted set of item tags that the user is likely to interact with, inferred from the joint analysis of user interaction history and profile information.

\begin{table}[t]
    \centering
    \caption{Criteria for Model-Generated Item Tags, where \textbf{Correct} indicates tags that meet all criteria, while \textbf{Incorrect} includes four categories: Weak Relevance, Low Consistency, Low Specificity, and Invalid Tag. ``All \correct\ $\rightarrow$ \correct'' means when all criteria are satisfied, the tag is considered correct, while ``Any \wrong\ $\rightarrow$ \wrong'' means if any criterion is violated, the tag is deemed incorrect.}
    \label{tab:reject_item_tag_criteria}
    \renewcommand{\arraystretch}{1.2} 
    \begin{tabularx}{\textwidth}{ccXX}
        \toprule
        \multicolumn{1}{c}{\textbf{Label}} & \multicolumn{1}{c}{\textbf{Evaluation Criteria }} & \multicolumn{1}{c}{\textbf{Example}} & \multicolumn{1}{c}{\textbf{Why \correct or \wrong}} \\
        \midrule
        \multirow{5}{*}[0ex]{\makecell{\textbf{Correct}\\(All \correct $\rightarrow$ \correct)}} 
            & \multirow{1}{*}[-0.7ex]{Strong Relevance} & \multirow{5}{*}{\makecell[l]{[Tag] \textit{Foldable Pet Case.} \\ {[}Interest] \textit{Cat Ownership.} \\ {[}Reason] \textit{User has a cat and} \\ \textit{focuses on portable storage.}}} & \multirow{2}{*}[-1ex]{\makecell[l]{Direct match to user need \\ and history. (Relevance\ \correct \& \\ Consistency\ \correct) \\Tag is specific and real. \\(Sepcificity\ \correct \& Validity\ \correct)}} \\
            & \multirow{1}{*}[-0.9ex]{High Consistency} &  &  \\
            & \multirow{1}{*}[-1.1ex]{High Specificity} &  &  \\
           & \multirow{1}{*}[-1.3ex]{Valid Tag} & &  \\
            & &  &  \\
        \midrule
        \multirow{4}{*}[-5ex]{\makecell{\textbf{Incorrect}\\(Any \wrong $\rightarrow$ \wrong)}} 
            & \multirow{1}{*}[0ex]{Weak Relevance} & \makecell[l]{[Tag] \textit{Embroidery bird-and} \\ \textit{-flower pattern silk pillowcase.}\\ {[}Interest] \textit{Skincare.}} & \multirow{1}{*}[0ex]{No direct link to skincare.} \\
            \cline{2-4}
            & \multirow{1}{*}[0ex]{Low Consistency} & \makecell[l]{[Tag] \textit{Calcium supplement} \\ \textit{powder for elderly health.} \\ {[}attribute age] \textit{20.} \\ {[}historical behaviors] \textit{None.}} & User is too young and no relevant history.  \\
            \cline{2-4}
             & \multirow{1}{*}[-1.5ex]{Low Specificity} & [Tag] \textit{Mountaineering outdoor sports equipment.} & \multirow{1}{*}[-1.5ex]{Too broad as a tag.} \\
            \cline{2-4}
            & Invalid Tag & [Tag] \textit{Smart coaster.} & Product does not exist. \\
        \bottomrule
    \end{tabularx}
\end{table}

\paragraph{Data Quality Control.}
To align $\mathcal{LLM}_{\text{IT}}$ with human consistency and enable it to function like a real shopping assistant, we also introduce multi-dimensional rejection sampling to achieve high-quality training sample filtering:
\begin{itemize}[itemsep=0.5em, topsep=0.5em, label=$\blacklozenge$]
    \item \textbf{Relevance:} Evaluates whether the generated tags are directly aligned with the user’s associated interests. This criterion measures the model’s capacity to genuinely understand and accurately predict user needs by assessing whether the tag matches the specified interest.
    \item \textbf{Consistency:} Assesses whether the item tag is generated with explicit reference to the user’s profile information and historical behavioral data. This criterion focuses on whether the model’s reasoning process incorporates authentic user context rather than fabricating or ignoring the given user information, ensuring that the generated tags are grounded in real user data.
    \item \textbf{Specificity:} Evaluates tag specificity to avoid generic term like ``fashion sports equipment'' that lead to imprecise product retrieval.
    \item \textbf{Validity:} Determines whether the predicted tags correspond to an actual existing product, preventing non-existent tag generation.
\end{itemize}

Based on the above multi-dimensional reject sampling criteria, we conduct strict quality control on the model-generated tags. Specifically, if a tag meets all criteria, it is labeled as a qualified sample for training; if any criterion is not satisfied, the tag is marked as an unqualified sample. Table~\ref{tab:reject_item_tag_criteria} shows the specific reject sampling criteria and examples. Through this approach, we ensure that $\mathcal{LLM}_{\text{IT}}$ can perform two-stage alignment on high-quality items that meet human evaluation standards, improving the accuracy and reliability of tag prediction.

\paragraph{Human Evaluation Experiments.}
To validate the effectiveness of our task alignment approach for item tag prediction, we conduct human evaluation on model-generated item tags according to the aforementioned criteria. A predicted tag is considered qualified only when it satisfies all evaluation standards. We compare four models: DeepSeek-R1, Qwen3-Base, Qwen3-SFT, and TBStars-SFT, where Qwen3-SFT and TBStars-SFT are full-parameter fine-tuned models based on our multi-stage task alignment framework.
As shown in Table~\ref{tab:human_eval_tag_prediction}, several key insights can be drawn from the results:

(1) Qwen3-Base achieves only 33.70\% pass rate, demonstrating substantial limitations when directly applying base LLMs to item tag prediction tasks. This poor performance indicates that foundation models lack sufficient domain knowledge and task-specific adaptation capabilities.

(2) Both aligned models, Qwen3-SFT (84.80\%) and TBStars-SFT (88.80\%), significantly outperform the DeepSeek-R1 (80.00\%). These results validate that our knowledge distillation from strong models and self-training evolution approach enables smaller-scale language models to progressively approach and eventually exceed the performance of  reasoning language models.

(3) TBStars-SFT achieves the highest performance at 88.80\% pass rate, substantially outperforming all other models. Beyond superior accuracy, the additional low-latency inference advantage of TBStars-SFT makes it particularly suitable for industrial recommender systems where both prediction quality and computational efficiency are necessary for practical deployment.

\begin{table}[t]
\centering
\caption{Human-evaluated pass rates for different models on item tag prediction task. The best performance is highlighted in \textbf{bold}.}
\label{tab:human_eval_tag_prediction}
\begin{tabular}{ccccc}
\toprule
\textbf{Model} & \textbf{DeepSeek-R1} & \textbf{Qwen3-Base} & \textbf{Qwen3-SFT} & \textbf{TBStars-SFT} \\
\midrule
\textbf{Pass Rate (\%)} & 80.00 & 33.70 & 84.80 & $\mathbf{88.80}$ \\
\bottomrule
\end{tabular}
\end{table}
\subsubsection{Incremental Learning}
\label{sec:Item_Tag_Prediction_Incremental_Learning}
To better adapt to dynamic user interests and data distribution shifts in online environments (such as seasonal changes), we adopt a bi-weekly \textit{\textbf{Incremental Learning (IL)}} method for updating the $\mathcal{LLM}_{\text{IT}}$.
During each update cycle, we select users' online interaction records (\textit{e.g.}, clicks, purchases) from the past 14 days as the data source for incremental training. 
However, real-world data presents two critical challenges: \textbf{(1) Substantial Noise}: e.g., accidental clicks or promotional artifacts, that misrepresent genuine preferences, and \textbf{(2) Inherent Imbalance}: dominant interest tags may skew model training—potentially degrading recommendation diversity, exacerbating filter bubbles, and reinforcing Matthew effects. 
To address these dual challenges, we design the following three-step process to handle online user behavior data:

\textbf{Step 1. Data Purification.} Following the data quality criteria for relevance and timeliness outlined in Section~\ref{sec:Item_Tag_Prediction_Task_Alignment}, we employ the QwQ-32B~\citep{qwq32b} as an automated judge for data cleaning. Specifically, for relevance, we analyze the consistency between user behaviors and their underlying interests, filtering out low-quality interaction records that do not align with user preferences. For timeliness, we focus on whether user behaviors satisfy seasonal requirements, \textit{i.e.}, whether the products are suitable for the current season or the upcoming season. This approach ensures high-quality training data by minimizing noise behaviors from random clicks and transient behaviors.

\textbf{Step 2. Interest Completion.} To construct structured training data, we need to map users' valid interaction behaviors to triplet outputs in the form of \textit{(Tag, Associated Interest Preference, Rationale)}. Specifically, we use QwQ-32B to perform deep reasoning based on given information (including user profiles, historical behaviors, and requirement prompts) to infer underlying interest preferences and the corresponding justifications that support user behaviors. Besides, since we can obtain real user interaction behaviors in this context, we directly use the item titles as tags. Through this approach, we can transform user behavioral data into structured data samples suitable for model training.

\textbf{Step 3. Data Balancing.} 
In this step, we design a two-stage data resampling strategy to address the inherent imbalance in online user behavior data.
Specifically, in the first stage, for each user, we first randomly select behavioral records corresponding to 80 item tags to ensure training data diversity and representativeness while improving training efficiency. In the second stage, we further utilize a pre-trained Tag-to-Cate model $\phi(\cdot)$ to convert these item tags into corresponding category labels (note that the number of categories is much smaller than the number of tags), and perform secondary sampling based on this, ensuring that the number of samples for each category is roughly equal (in our experimental setup, we sample at most 2 samples per category) to achieve data balance.

Through the above process, we ultimately obtain high-quality, diversified incremental online training data. This incremental learning strategy not only helps the $\mathcal{LLM}_{\text{IT}}$ learn the dynamic changes in users' latest preferences and product knowledge, but also effectively improves the model's generalization capability and recommendation accuracy, avoiding repeated recommendations of duplicate and outdated products, and achieving weekly optimization for industrial RS.

\begin{table}[h]
\centering
\caption{Performance comparison of tag prediction accuracy before and after using incremental learning (IL). The best performance is highlighted in \textbf{bold}.}
\label{tab:incremental_learning_exp}
\begin{tabular}{ccccc}
\toprule
\textbf{Model} & \textbf{TBStars-SFT (w/o IL)} & \textbf{TBStars-SFT (w/ IL)} \\
\midrule
\textbf{HR@30} & 0.3671 & $\mathbf{0.3776}~~(\mathbf{+ 1.05\%})$ \\
\bottomrule
\end{tabular}
\end{table}

\paragraph{Incremental Learning Effectiveness Evaluation}
To validate the effectiveness of incremental learning, we leverage real online user interaction behaviors for verification. Specifically, we utilize the $\mathcal{LLM}_{\text{IT}}$ to predict 30 item tags for each user's historical sequence according to Eq.~\eqref{eq:it}, and employ the Tag-to-Gate model $\phi(\cdot)$ to convert these tags $\mathcal{T}_{u}=\{T_{u,1},T_{u,2},\dots,T_{u,30}\}$ into specific predefined product category $\mathcal{C}_u^\text{pred} = \{C_{u,1},C_{u,2},\dots,C_{u,30}\}$. We design the HR@30 metric to evaluate item tag prediction accuracy, which is formulated as:
\begin{equation*}
\text{HR@30} = \frac{1}{|\mathcal{U}|} \sum_{u \in \mathcal{U}} \mathbb{I}(C_{u}^\text{gt} \in \mathcal{C}_u^{\text{pred}}),
\end{equation*}
where $\mathcal{U}$ represents the set of test users, $C_{u}^\text{gt}$ denotes the product category of user $u$'s actual next interacted item, $\mathcal{C}_u^{\text{pred}}$ represents the set of predicted categories converted from the 30 predicted item tags for user $u$, and $\mathbb{I}(\cdot)$ is the indicator function that returns 1 if the predicted category set contains the true next interaction category, and 0 otherwise.

To demonstrate the practical value of our incremental learning approach, we conduct a comparative analysis using TBStars-SFT, the foundation recommendation model currently deployed in our online system. We evaluate the prediction accuracy before and after applying incremental training on real-world data.
As presented in Table, we observe that the model fine-tuned with cleaned and balanced datasets from online interactions achieves a notable $1.05\%$ improvement in HR@30 compared to the baseline model without incremental learning. This improvement is particularly significant considering the scale and complexity of real-world recommendation scenarios, where even marginal gains can translate to substantial business impact. The results validate the effectiveness of our incremental learning strategy in adapting to evolving user preferences and emerging product trends. 
\subsection{Item Retrieval}
\label{sec:Item_Retrieval}
\begin{figure}[t]
    \centering
    \includegraphics[width=0.8\textwidth]{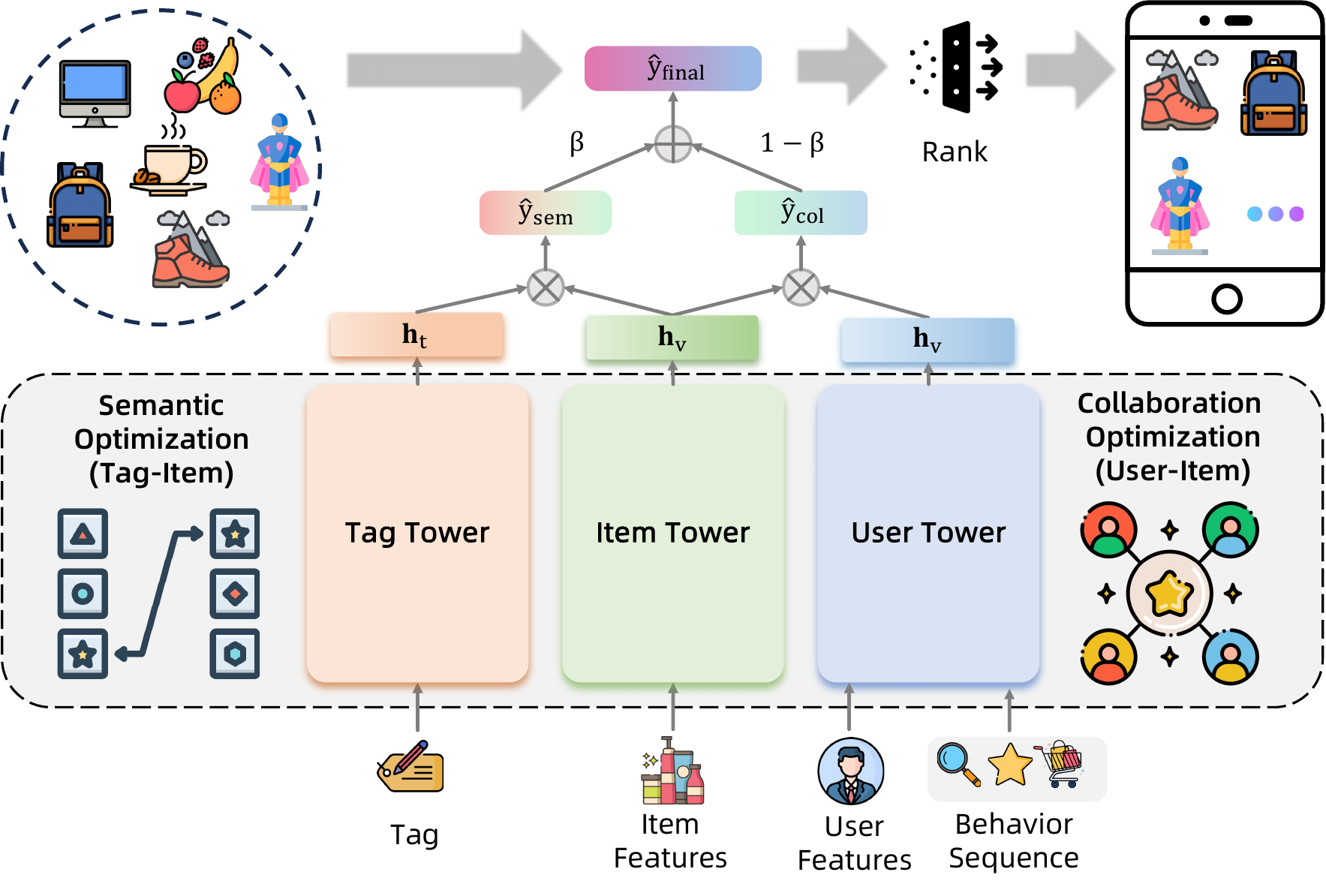}
    \caption{User-Item-Tag Retrieval Framework. The framework jointly optimizes tag tower and item tower for semantic enhancement, while combining user tower and item tower for collaborative optimization. Based-on collaborative-semantic relevance scoring, retrieved items are aligned with user interests at the source level, with filtered items subsequently fed into the ranking stage.}
    \label{fig:tag}
\end{figure}

While the LLM-generated item tags provide rich semantic understanding of user preferences, a critical challenge emerges: these abstract semantic representations cannot be directly mapped to specific products within the target domain. The gap between high-level semantic concepts and concrete item characteristics necessitates an effective bridge mechanism for practical recommendation deployment. To address this challenge, we introduce a tag-aware method that seamlessly connects semantic understanding with item retrieval. Furthermore, recognizing that collaborative knowledge derived from user-item interactions contains valuable behavioral patterns complementary to semantic signals, we integrate collaborative filtering mechanisms to enhance retrieval effectiveness. This leads to a unified \textbf{User-Item-Tag Retrieval Framework} that synergistically combines semantic reasoning capabilities with collaborative behavioral insights, ultimately improving both the accuracy and efficiency of online recommender systems.

In the following sections, we present the overall architecture of user-item-tag retrieval framework (Section~\ref{sec:Item_Retrieval_Architecture}), the collaborative-semantic enhanced optimization algorithm (Section~\ref{sec:Item_Retrieval_Optimization}) and the online inference methodology (Section~\ref{sec:Online_Inference}).

\subsubsection{Overall Architecture}
\label{sec:Item_Retrieval_Architecture}
The overall architecture of TAR is illustrated in Figure~\ref{fig:tag}.
The framework consists of three parallel towers: the Item Tower, User Tower, and Tag Tower. We introduce the details of each tower as follows:

\textbf{Item Tower:} Given an item $v$, we define its feature set as $\mathcal{F}_v = \{\mathcal{F}_v^{\text{sparse}}, \mathcal{F}_v^{\text{dense}}\}$, where $\mathcal{F}_v^{\text{sparse}} = \{item\ id, category, brand, ...\}$ represents sparse categorical features and $\mathcal{F}_v^{\text{dense}} = \{price, sales, ...\}$ denotes continuous numerical features.
The feature embedding layer transforms both sparse and discretized dense features into uniform dense vectors:
\begin{align*}
\mathbf{e}_j^{\{\text{sparse}\}} &= \text{EMB}(f_j^{\{\text{sparse}\}}), \quad f_j^{\{\text{sparse}\}} \in \mathcal{F}_v^{\{\text{sparse}\}} \\
\mathbf{e}_k^{\{\text{dense}\}} &= \text{EMB}(\text{Discretize}(f_k^{\{\text{dense}\}})), \quad f_k^{\{\text{dense}\}} \in \mathcal{F}_v^{\{\text{dense}\}}
\end{align*}
where $\text{EMB}(\cdot)$ denotes the embedding operation, and $\text{Discretize}(\cdot)$ converts continuous features into discrete representations.
The item tower then applies a Deep Neural Network (DNN) to learn the item representation:
\begin{equation*}
\mathbf{h}_v = \text{DNN}_{\text{item}}([\mathbf{e}_1^{\{\text{sparse}\}}; \mathbf{e}_2^{\{\text{sparse}\}}; \dots; \mathbf{e}_1^{\{\text{dense}\}}; \mathbf{e}_2^{\{\text{dense}\}}; ...])
\end{equation*}

\textbf{User Tower:} The user tower captures user preferences through multi-behavioral sequence modeling. For user $u$, the input features include user ID and multi-behavior interaction sequences: $\mathcal{F}_u = \{user\ id, \mathcal{S}_u^{\text{click}}, \mathcal{S}_u^{\text{purchase}}, ...\}$, where $\mathcal{S}_u^{\text{behavior}}$ represents the chronological sequence of user interactions for a specific behavior type.
For each behavior sequence, we apply mean pooling over the item representations to obtain the sequence representation $\mathbf{s}_u^{\text{behavior}}$.

The user representation is then computed by concatenating the user ID embedding with the pooled sequence representations:
\begin{equation*}
\mathbf{h}_u = \text{DNN}_{\text{user}}([\text{EMB}(id_u); \mathbf{s}_u^{\text{click}}; \mathbf{s}_u^{\text{purchase}}; ...])
\end{equation*}

\textbf{Tag Tower:} The tag tower transforms the item tag $T$ into dense representations:
\begin{equation*}
\mathbf{h}_t = \text{DNN}_{\text{tag}}(\text{MEAN}([\text{EMB}(w_1); \text{EMB}(w_2); ...; \text{EMB}(w_n)]))
\end{equation*}
where $w_i$ represents the $i$-th token in the tokenized tag sequence $T = [w_1, w_2, ..., w_n]$, and $\text{MEAN}(\cdot)$ denotes the mean pooling operation.

Our framework generates two complementary prediction scores:
\begin{align*}
\hat{y}_{\text{col}} &= \mathbf{h}_u^T \mathbf{h}_v \quad \text{(Collaborative Score)},\\
\hat{y}_{\text{sem}} &= \mathbf{h}_t^T \mathbf{h}_v \quad \text{(Semantic Score)}.
\end{align*}
The score $\hat{y}_{\text{col}}$ captures behavioral collaborative patterns through user-item interaction modeling, while the score $\hat{y}_{\text{sem}}$ leverages tag-item semantic relevance to understand preference reasoning.

\subsubsection{Optimization}
\label{sec:Item_Retrieval_Optimization}
In this section, we introduce the optimization objective functions for user-item collaborative modeling and tag-item semantic modeling, respectively.
\paragraph{Collaborative Optimization.}
The collaborative optimization objective is to maximize the likelihood of positive user-item interactions while minimizing the likelihood of negative interactions. 
Specifically, we treat items clicked by users as positive samples and unclicked items as negative samples, employing negative sampling from the latter to perform contrastive learning-based optimization.
The optimization objective is formulated as follows:
\begin{equation*}
\mathcal{L}_{\text{col}} = -\sum_{(u,v) \in \mathcal{D}} \log \frac{\exp(\mathbf{h}_u^T \mathbf{h}_v)}{\exp(\mathbf{h}_u^T \mathbf{h}_v) + \sum_{v' \in \mathcal{V}^-} \exp(\mathbf{h}_u^T \mathbf{h}_{v'})}
\end{equation*}
where $\mathcal{D}$ represents the set of positive user-item pairs, and $\mathcal{V}^-$ denotes the set of sampled negative items for each user.

\paragraph{Semantic Optimization.}
In contrast to collaborative optimization, semantic optimization aims to maximize the semantic relevance between tags generated based on predicted user preferences and items. We adopt a similar contrastive learning-based optimization, where the positive samples are the tags of clicked items, and the negative samples are randomly sampled tags from other items:
\begin{equation*}
\mathcal{L}_{\text{tag}} = -\sum_{(t,v) \in \mathcal{D}} \log \frac{\exp(\mathbf{h}_t^T \mathbf{h}_v)}{\exp(\mathbf{h}_t^T \mathbf{h}_v) + \sum_{v' \in \mathcal{V}^-} \exp(\mathbf{h}_t^T \mathbf{h}_{v'})}
\end{equation*}
where $\mathcal{V}^-$ denotes sampled negative items for each tag.

Furthermore, to prevent overfitting to descriptive tag features, we introduce a category contrastive loss function to enhance semantic discrimination within item categories. Specifically, for each original tag-item pair $(t,v)$ from $\mathcal{D}$, we sample items from the same category as $v$ to serve as positive samples and items from different categories as negative samples:
\begin{equation*}
\mathcal{L}_{\text{cate}} = -\sum_{(t,v) \in \mathcal{D}} \sum_{v^+ \in \mathcal{C}^+_v} \log \frac{\exp(\mathbf{h}_t^T \mathbf{h}_{v^+})}{\exp(\mathbf{h}_t^T \mathbf{h}_{v^+}) + \sum_{v^- \in \mathcal{C}^-_v} \exp(\mathbf{h}_t^T \mathbf{h}_{v^-})}
\end{equation*}
where $\mathcal{C}^+_v$ represents the set of sampled items from the same category as item $v$, and $\mathcal{C}^-_v$ represents the set of sampled negative items from different categories than item $v$. This category-aware contrastive learning encourages the model to learn fine-grained semantic distinctions within categories while maintaining clear boundaries across different categories.

The final optimization objective of TAR is formulated as:
\begin{equation*}
\mathcal{L}_{\text{TAR}} = \mathcal{L}_{\text{col}} + \underbrace{\alpha\mathcal{L}_{\text{tag}} + (1 - \alpha) \mathcal{L}_{\text{cate}}}_{\mathcal{L}_{\text{sem}}},
\end{equation*}
where $\alpha$ is a hyperparameter that balances the contributions of tag and category contrastive losses. We set $\alpha = 0.5$ in our experiments, indicating equal importance for both losses.

\subsubsection{Online Inference}
\label{sec:Online_Inference}
During the inference phase, we dynamically fuse the outputs of the user tower and tag tower to achieve controllable recommendations with collaborative-semantic relevance.
Specifically, we first compute the output vectors $\mathbf{h}_u$ and $\mathbf{h}_t$ from the user tower and tag tower (where tags are predicted by $\mathcal{LLM}_{\text{IT}}$), and then perform weighted fusion:
\begin{equation*}
\mathbf{h}_{\text{fuse}} = \beta \mathbf{h}_u + (1 - \beta) \mathbf{h}_t
\end{equation*}
where $\beta$ is a hyperparameter controlling the fusion ratio between user tower and tag tower outputs. This fused representation is used for item retrieval from the candidate pool. Essentially, it is equivalent to computing the final matching score as a weighted sum of the collaborative and semantic scores:
\begin{equation*}
\hat{y}_{\text{final}} = \beta \hat{y}_{\text{col}} + (1 - \beta) \hat{y}_{\text{sem}},
\end{equation*}
where the collaborative and semantic signals are balanced according to the fusion weight. The resulting matching scores are then utilized in the downstream recommendation pipeline.
This dynamic fusion mechanism enables flexible control over the balance between collaborative filtering signals and semantic understanding, allowing the system to adapt to different recommendation scenarios while maintaining both behavioral relevance and semantic coherence.


\subsection{Personalized Explanation Generation}
\label{sec:Explanation_Generation}
Beyond enhancing the matching between candidate items and user interests, RecGPT also introduces a recommendation explanation generation module to further elevate user experience in recommender systems. This module generates personalized explanations for recommended items, helping users better understand recommendation outputs by answering the fundamental question: ``\textit{why is this product recommended to me}''.
Below, we detail the \textbf{Task Alignment} for the Recommendation-Explanation LLM $\mathcal{LLM}_{\text{RE}}$ and the \textbf{Offline Production} strategy to meet online low-latency requirements.

\begin{figure}[t]
    \centering
    \includegraphics[width=\textwidth]{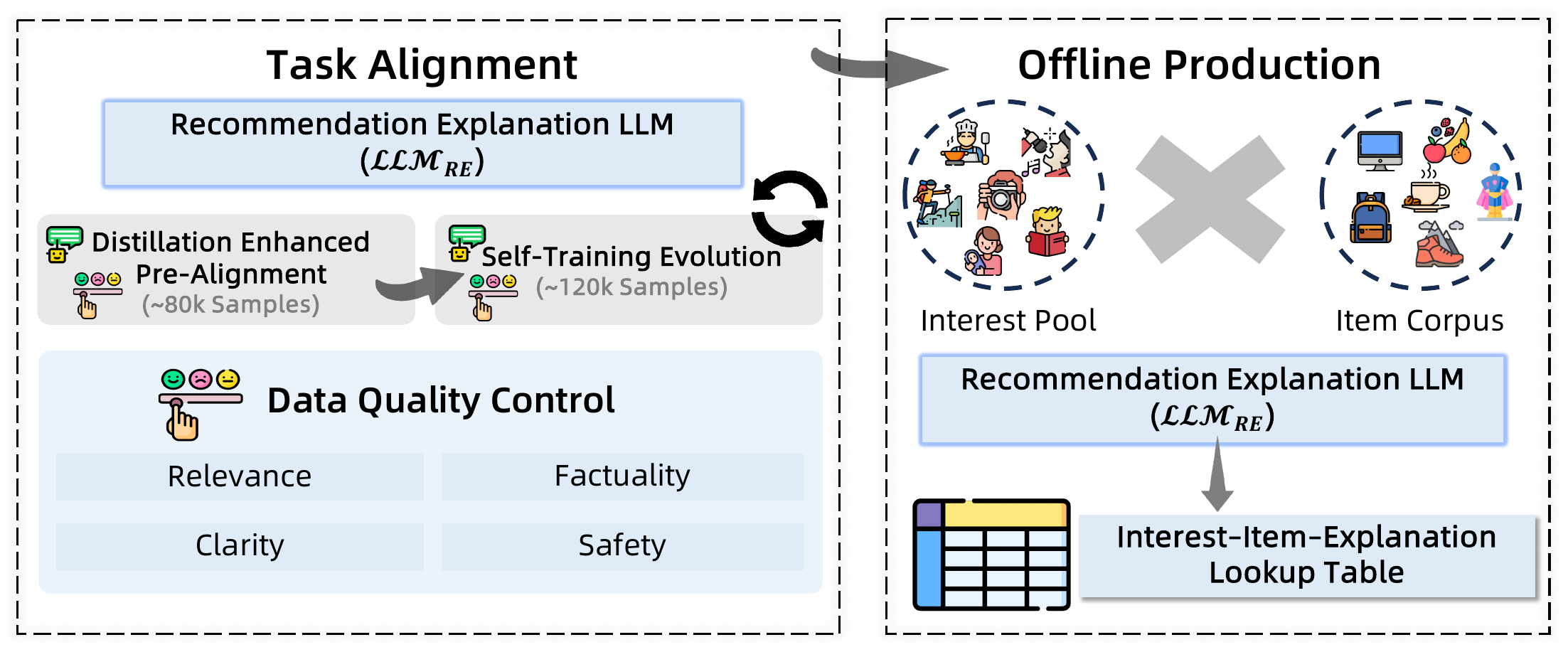}
    \caption{Illustration of the recommendation explanation generation task. The left figure demonstrates the task alignment process and data quality control protocols, while the right figure shows the offline production of interest-item-explanation tables using the recommendation-explanation LLM.}
    \label{fig:explanations}
\end{figure}

\subsubsection{Task Alignment for Recommendation Explanation Generation}
\label{sec:Explanation_Generation_Task_Alignment}
Similar to the item tag prediction tasks, we adapt large language models for recommendation explanation generation through two-stage training. We first pre-train the model using reasoning-enhanced teacher datasets generated by DeepSeek-R1 (\textbf{Reasoning-Enhanced Pre-Alignment}), followed by training on self-generated data that are subject to rigorous quality control through human or LLM-Judge filtering (\textbf{Self-Training Evolution}), ultimately achieving human-aligned explanation generation performance.
In this section, we focus on the \textbf{Prompt Engineering} and \textbf{Data Quality Control} protocols specifically designed for this task.
Additionally, we provide detailed \textbf{Human Evaluation Experiments} to validate the effectiveness of our alignment approach.

\paragraph{Prompt Engineering.}

Given user interest sets and relevant recommended item information (such as item tags, titles), we instruct $\mathcal{LLM}_{\text{RE}}$ to execute the following two steps to generate reasonable recommendation explanations:
\begin{enumerate}[itemsep=0.5em, topsep=0.em]
    \item \textbf{Context Understanding.} Analyze the given input information to understand user interests and item characteristics.
    \item \textbf{Explanation Generation.} Based on the above analysis, if reasonable correlations exist between recommended products and user interests, generate conversational phrases that present these connections while maintaining an approachable tone; otherwise, generate recommendation explanations primarily based on the product's inherent qualities.
\end{enumerate}
Through these steps, the model $\mathcal{LLM}_{\text{RE}}$ can generate personalized recommendation explanations based on user interests and item information, helping users understand recommendation results and enhancing user experience.
The simplified prompt template is shown as Prompt~\ref{box:exp-generation}, where the placeholders \textcolor{BlueViolet}{\textbf{\{User Interest\}}}, \textcolor{BlueViolet}{\textbf{\{Date Information\}}}, and \textcolor{BlueViolet}{\textbf{\{Item Information\}}} are instantiated with specific user interests, current date, and recommended item information, respectively. The placeholders \textcolor{Maroon}{\textbf{\{Context Understanding\}}}, and \textcolor{Maroon}{\textbf{\{Explanation Generation\}}} are populated with the two reasoning steps mentioned above, designed to leverage the CoT-based reasoning capabilities of LLMs to generate reasonable explanations, while \textcolor{Maroon}{\textbf{\{Recommendation Principles\}}} and \textcolor{Maroon}{\textbf{\{Strict Prohibitions\}}} are filled with pre-configured requirements and constraints.
Due to space constraints, we provide details about the full prompt template in the Appendix~\ref{box:app-exp-generation}.

\begin{tcolorbox}[promptbox=exp_colframe/exp_colback, title={Recommendation Explanation Generation Prompt Template},label=box:exp-generation,breakable]
\colorbox{green!20}{\textbf{\# Role}}\\
Generate personalized recommendation explanations based on user profiles and recommended items. The explanations must satisfy the following requirements.\\
\colorbox{blue!20}{\textbf{\# Input}}\\
\textbf{User Interest}: \textcolor{BlueViolet}{\textbf{\{User Interest\}}}\\
\textbf{Current Date}: \textcolor{BlueViolet}{\textbf{\{Date Information\}}}\\
\textbf{Item Information}: \textcolor{BlueViolet}{\textbf{\{Item Information\}}} \\
\colorbox{orange!20}{\textbf{\# Core Reasoning Steps}}\\
\textcolor{Maroon}{\textbf{\{Context Understanding\}}} | \textcolor{Maroon}{\textbf{\{Explanation Generation\}}}\\
\colorbox{red!20}{\textbf{\# Mandatory Requirements}}\\
\textbf{**Recommendation Principles (\correct)**}\\
\textcolor{Maroon}{\textbf{\{Recommendation Principles\}}}\\
\textbf{**Strict Prohibitions (\wrong)**}\\
\textcolor{Maroon}{\textbf{\{Strict Prohibitions\}}}\\
\colorbox{yellow!20}{\textbf{\# Output Format}}\\
\textit{(Detailed output format requirements)}
\end{tcolorbox}

Based on the aforementioned prompt engineering design, we formalize the explanation generation process as follows:
Given user interest $\mathcal{I}_u$ and item information $\text{Info}_v$, we utilize the recommendation explanation LLM to generate personalized explanations for the recommended items:
\begin{equation}
E_u = \mathcal{LLM}_{\text{RE}}(\mathcal{I}_u, \text{Info}_v | \mathcal{P}_{RE}),
\end{equation}
where $E_u$ represents the generated explanation for user $u$, $\mathcal{LLM}_{\text{RE}}(\cdot)$ denotes the recommendation explanation model, $\mathcal{I}_u$ denotes the user's interests, $\text{Info}_v$ contains the relevant information about item $v$ (\textit{e.g.}, item tags, titles), and $\mathcal{P}_{RE}$ represents the prompt template.

\begin{table}[h]
    \centering
    \caption{Criteria for rejecting model-generated explanations. In this example, the user's interest is \textit{outdoor travel}, and the product is a \textit{backpack}.}
    \label{tab:reject_explanation_criteria}
    \renewcommand{\arraystretch}{1.2} 
    \begin{tabularx}{\textwidth}{ccXX}
        \toprule
        \multicolumn{1}{c}{\textbf{Label}} & \multicolumn{1}{c}{\textbf{Evaluation Criteria}} & \multicolumn{1}{c}{\textbf{Example}} & \multicolumn{1}{c}{\textbf{Why \correct or \wrong}} \\
        \midrule
        
        \multirowcell{4}{\textbf{Accept}\\ (Both \correct $\rightarrow$ \correct)} 
        & Strong \textbf{Relevance} 
        & \multirow{4}{*}{\makecell[l]{\textit{Roam mountains rivers}\\ \textit{backpack journey}\\ \textit{companion.}}} 
        & \multirow{4}{*}{\makecell[l]{Aligning interests with use.\\ (Relevance \& Factuality \correct)\\ Brief, poetic, no privacy\\ leaked. (Clarity \& Safety \correct)}} \\
        & Verified \textbf{Factuality} & & \\
        & Full \textbf{Clarity} & & \\
        & Proven \textbf{Safety} & & \\
        
        \midrule[0.8pt]
        
        \multirowcell{8}{\textbf{Reject}\\ (Any \wrong $\rightarrow$ \wrong)} 
        & Weak \textbf{Relevance} 
        & \textit{Office backpack document carry convenience.} 
        & Ignores user's outdoor interest. \\
        \cmidrule(lr){2-4}
        
        & Unverified \textbf{Factuality} 
        & \textit{Bag cutproof fireproof lasts forever.} 
        & Exaggerated false claims. \\
        \cmidrule(lr){2-4}
        
        & Limited \textbf{Clarity} 
        & \textit{Bag bag good bag buy bag good.} 
        & Repetitive nonsense. \\
        \cmidrule(lr){2-4}
        
        & Unproven \textbf{Safety} 
        & \textit{MsZhang time-limited offer bag quick buy.} 
        & Privacy leak + hard sell. \\
        
        \bottomrule
    \end{tabularx}
\end{table}

\paragraph{Data Quality Control.}
To ensure the model's instruction-following capability, we also introduce multi-dimensional rejection sampling to achieve high-quality training sample filtering:
\begin{itemize}[topsep=0em, itemsep=0.5em, label=$\blacklozenge$]
    \item \textbf{Relevance}: Alignment between the explanation and both the characteristics of the recommended item and the user's interests.
    \item \textbf{Factuality}: Accuracy of the explanation in reflecting the item's actual features and functionality.
    \item \textbf{Clarity}: Quality of text fluency, grammatical correctness, and stylistic expression.
    \item \textbf{Safety}: Absence of sensitive or personally identifiable information in the generated content.
\end{itemize}

Positive and negative examples of the above criteria are shown in Table~\ref{tab:reject_explanation_criteria}. We consider samples with generated explanations that meet the above criteria as qualified samples, while those that do not meet the criteria are regarded as unqualified samples. During the recommendation explanation LLM alignment process, we employ both human evaluation and LLM auto-evaluation to filter training samples, improving the model's instruction-following ability and generation quality.

\begin{table}[h]
\centering
\caption{Human-evaluated pass rates for different models on recommendation explanation generation task. The best performance is highlighted in \textbf{bold}.}
\label{tab:human_eval_explanation_generation}
\begin{tabular}{ccccc}
\toprule
\textbf{Model} & \textbf{DeepSeek-R1} & \textbf{Qwen3-Base} & \textbf{Qwen3-SFT} \\
\midrule
\textbf{Pass Rate (\%)} & 92.7 & 30.0 & \textbf{95.8} \\
\bottomrule
\end{tabular}
\end{table}

\paragraph{Human Evaluation Experiments.}
To validate the effectiveness of our task alignment approach for recommendation explanation generation, we conduct human evaluation on model-generated explanations according to the multi-dimensional criteria outlined above. An explanation is considered qualified only when it satisfies all evaluation standards including relevance, factuality, clarity, and safety. We compare three models: DeepSeek-R1, Qwen3-Base, and Qwen3-SFT, where Qwen3-SFT represents our multi-stage aligned LLMs.

As shown in Table~\ref{tab:human_eval_explanation_generation}, our experimental analysis reveals several important findings:

(1) Qwen3-Base demonstrates insufficient performance in generating high-quality recommendation explanations that meet industry standards, with a pass rate of only 30\%. This limitation stems from the lack of domain-specific knowledge and task-oriented instruction-following capabilities required for personalized explanation generation in recommendation scenarios.

(2) DeepSeek-R1 achieves superior performance compared to Qwen3-Base with 92.7\% pass rate, benefiting from its larger parameter scale and enhanced reasoning capabilities. The model's deep thinking abilities enable better understanding of user-item relationships and generation of more coherent explanations that align with user interests and item characteristics.

(3) Our aligned Qwen3-SFT model demonstrates substantial improvement in adapting to multi-dimensional explanation generation requirements, achieving the highest pass rate of 95.8\%. Through training on carefully curated high-quality recommendation explanation samples via reasoning-enhanced pre-alignment and self-training evolution, the model effectively learns to generate explanations that satisfy relevance, factuality, clarity, and safety criteria simultaneously. This comprehensive alignment makes it well-suited for online deployment requirements where both explanation quality and computational efficiency are essential for large-scale recommender systems.

\subsubsection{Offline Production}
\label{sec:Explanation_Generation_Offline_Production}
Due to the excessive computational overhead of generating recommendation explanations for each user--item pair in real--time online scenarios, it becomes challenging to meet the low--latency requirements of industrial recommender systems. To address this issue, we design an interest--based offline explanation production method.

Specifically, we start from the user interest set and leverage the collected tag--interest association pairs (see Section~\ref{sec:Item_Tag_Prediction_Task_Alignment}). We utilize a pretrained Tag--to--Cate model $\phi(\cdot)$ to map the predicted item tag $T$ to specific item categories, which can be formalized as:
\begin{equation*}
C \leftarrow \phi(T),
\end{equation*}
where $C$ represents the mapped item category of the item tag.
Since each item can be mapped to its corresponding item category, we can establish associations between user interests and individual items through their shared categories. This creates pairing relationships between user interests and specific items within the same category. Importantly, this approach generates only matched \textit{interest--item} pairs rather than exhaustive \textit{user--item} combinations, significantly reducing the target scope.

Based on this framework, we perform offline explanation generation for all matched interest--item pairs, creating a comprehensive \textbf{Interest--Item--Explanation Lookup Table}. During online recommendation, we efficiently retrieve the corresponding explanations from this precomputed table by matching the currently recommended items with the user's interest set. This approach enables real--time explanation delivery while dramatically reducing computational overhead compared to generating explanations for all possible user--item combinations.


\section{Human-LLM Cooperative Judge}
\label{sec:Human_LLM_Cooperation}
To ensure that large language models meet human subjective expectations in recommendation generation tasks (\textit{i.e.}, user interest mining, item tag prediction, and recommendation explanation generation), we manually curate training samples generated by DeepSeek-R1 or our self-trained models to align with human standards. 
However, scaling up manual evaluation through crowdsourced annotation is impractical in real-world industrial environments due to prohibitive costs and lengthy development cycles. 
Inspired by the excellent performance of \textbf{\textit{LLM-as-a-Judge}} approaches across various natural language understanding and generation tasks~\citep{gu2024survey,zheng2023judging,chen2024mllm,tang2025hf4rec}, we adopt this paradigm by leveraging LLMs as intelligent judges to achieve automated evaluation, aiming to reduce evaluation costs and improve efficiency.

However, we empirically find two critical challenges that hinder the effectiveness of LLM-Judges:
\begin{itemize}[topsep=0em, itemsep=0.5em]
    \item \textbf{Cognitive Bias:} Unlike straightforward evaluation tasks like harmfulness assessment, recommendation systems require understanding complex user behaviors, product characteristics, and operational strategies. This demands domain-specific knowledge and contextual awareness beyond basic reasoning capabilities. Native LLMs often exhibit cognitive biases due to knowledge limitations and pre-training biases~\citep{ye2024justice,schroeder2024can,son2024llm, dai2024bias}, compromising their evaluation reliability.
    \item \textbf{Temporal Misalignment:} The dynamic nature of recommendation ecosystems creates a fundamental mismatch between static LLM judges and evolving real-world conditions. This temporal discrepancy manifests through three critical dimensions: 
    \begin{itemize}
        \item \textit{Evolving User Behavior Patterns} -- emerging interaction trends and shifting preference distributions that deviate from historical training data.
        \item \textit{Dynamic Item Characteristics} -- introduction of new product categories, features, and attributes that were not present during judge training.
        \item \textit{Updated Evaluation Criteria} -- evolving business strategies, market expectations, and quality standards that continuously redefine the evaluation criteria. 
    \end{itemize}
    The cumulative effect of these temporal dynamics progressively undermines the evaluation capabilities of static LLM judges, introducing systematic biases for different generation tasks.
\end{itemize}
To address these issues, we propose a Human-LLM Cooperative Judge System. The core idea is to enhance task-specific evaluation capabilities through collaborative cooperation between human experts and LLM-Judge, while integrating human-in-the-loop supervision that monitors performance milestones and triggers realignment with evolving data distributions and task requirements when needed. In the following sections, we will provide detailed introductions to the two key components: \textbf{LLM-as-a-Judge} (Section~\ref{sec:LLM_as_a_Judge}) and \textbf{Human-in-the-Loop} (Section~\ref{sec:Human_in_the_Loop}).

\begin{figure}[t]
    \centering
    \includegraphics[width=\textwidth]{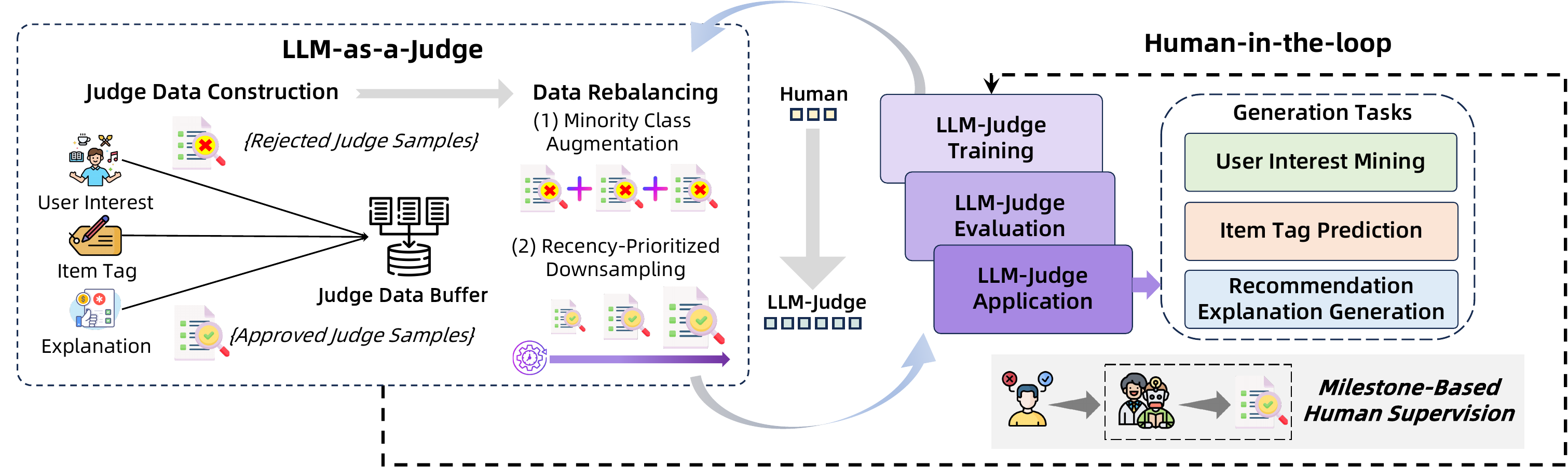}
    \caption{Human-LLM Cooperative  Judge System. Multi-round human judgment data from three generative tasks (user interest mining, item tag prediction, recommendation explanation generation) are collected into a Judge Data Buffer, with data balancing through minority class augmentation and recency-prioritized downsampling. The LLM-Judge is trained and deployed upon reaching accuracy thresholds, complemented by periodic human performance evaluation. This judge system facilitates a gradual transition from human curation to LLM-Human Cooperative curation.} 
    \label{fig:judge}
\end{figure}

\subsection{LLM-as-a-Judge}
\label{sec:LLM_as_a_Judge}
To enhance the alignment between LLM-based Judges and human evaluators in recommendation generation tasks, we develop a human-annotated evaluation dataset for LLM instruction fine-tuning.

\paragraph{Dataset Construction.}
Specifically, we first categorize the evaluation tasks across different generation tasks and assessment criteria into the following two types:
\begin{itemize}[itemsep=0.5em,topsep=0.em]
    \item \textbf{Binary Classification Evaluation}, \textit{e.g.}, in item tag prediction task, we employ a binary \{Yes, No\} evaluation scheme for ``\textit{Relevance}'', determining whether the tags are relevant to user interests.
    \item \textbf{Multi-level Evaluation}, \textit{e.g.}, in recommendation explanation generation task, we adopt a multi-level evaluation scheme \{Excellent, Good, Bad\} for ``\textit{Truthfulness}'', assessing how well the generated recommendation explanations align with factual information.
\end{itemize}
Furthermore, we collect judge training data from the following sources:
\begin{itemize}[itemsep=0.5em,topsep=0.em]
    \item \textbf{Pre-alignment Data:} Reasoning-enhanced data generated by DeepSeek-R1 during the pre-alignment phase.
    \item \textbf{Self-training Data:} Self-Generated samples produced from the task-specific LLM across multiple iterative rounds during the self-training phase.
\end{itemize}
We conduct human annotation on data from both sources for quality assessment according to different evaluation criteria specific to their respective tasks. The annotated samples and results are stored in a \textit{\textbf{Judge Data Buffer}}, which is subsequently used to fine-tune corresponding LLM-Judges. 

\paragraph{Data Rebalancing Strategy.}
However, in practice, we observed severe class imbalance in the collected judge training data, leading to significant \textit{Majority Class Bias} in the trained judge models. Under the \textit{Empirical Risk Minimization (ERM)} principle~\citep{johnson2019survey}, models predominantly learn from majority classes during training, thereby neglecting minority class characteristics and compromising model generalization and evaluation accuracy.
To address this challenge, we design a data rebalancing strategy for judge model training, comprising the following steps:

\textbf{(1) Minority Class Augmentation:} For underrepresented classes, we cumulatively utilize samples from multiple previous rounds of human annotation to augment data for these categories.

\textbf{(2) Recency-Prioritized Downsampling:} For dominant classes, we employ a temporal decay-based downsampling strategy that prioritizes the most recent evaluation samples while gradually incorporating earlier samples, effectively balancing sample quantities across different classes.

We empirically find that our proposed resampling strategy effectively improve the evaluation performance, particularly in evaluation accuracy for minority classes. This approach effectively enhances model generalization capabilities and prevents learning collapse. Note that our current approach primarily focuses on \textit{data-level} balancing strategies. Future work will explore \textit{model-level} balancing techniques, such as cost-sensitive learning~\citep{elkan2001foundations,fernandez2018cost}.
\subsection{Human-in-the-Loop}
\label{sec:Human_in_the_Loop}
While LLM-as-a-Judge systems demonstrate significant advantages in cost-effectiveness and evaluation efficiency, their reliability faces critical challenges due to dynamic data distribution shifts. LLM-based Judges become not only increasingly unreliable in quality assessmen but also struggle to adapt to evolving evaluation standards when encountering emerging user behavior patterns or novel product characteristics.

To address these limitations, we propose a \textit{Milestone-Based Human Supervision} framework that integrates human-in-the-loop validation. Specifically, during major version updates: \textit{First}, we collect expert annotations on recent generation samples; \textit{Second}, we perform systematic comparisons between LLM Judge evaluations and human assessments. When detecting substantial performance degradation, we conduct continuous training through targeted fine-tuning of the LLM-Judge using newly annotated data. This dual approach ensures sustained alignment with evolving data distributions while maintaining operational efficiency.

By combining \textit{automated LLM-as-a-judge evaluation} with \textit{strategic human-in-the-loop oversight}, we establish a robust human-LLM cooperative judgment system. This hybrid framework enables reliable large-scale data curation and model performance monitoring, achieving an optimal balance between evaluation accuracy and operational efficiency.


\section{Evaluation}
\label{sec:Evaluation}
In this section, we first introduce the experimental setup of RecGPT, including user group selection of online serving, training infrastructure, and implementation details (\textbf{Section~\ref{sec:Evaluation_Setup}}). We then present the overall performance of RecGPT in online A/B testing, analyzing its impact on users, merchants, and the platform (\textbf{Section~\ref{sec:Online_AB_Test}}). We examine the consistency between LLM-as-a-Judge and human evaluators in different recommendation generation tasks (\textbf{Section~\ref{sec:LLM_as_a_Judge_Evaluation}}). 
Additionally, we demonstrate real-world case studies (\textbf{Section~\ref{sec:Case_Studies}}) and conduct user surveys (\textbf{Section~\ref{sec:User_Experience_Investigation}}) with online users to capture user feedback and reflect the changes brought by RecGPT.

\subsection{Evaluation Setup}
\label{sec:Evaluation_Setup}
\paragraph{User Group Selection.} We conducted a one-month online A/B experiment by deploying RecGPT to the ``Guess What You Like'' (\textbf{Guess}) scenario on Taobao's homepage. The experiment targeted the top one-third of active users, with both control and experimental groups each allocated 1\% of the traffic. Users in the experimental group received recommendations generated from RecGPT system, while those in the control group continued using the existing base recommender system.
\paragraph{Infrastructure.} To optimize computational efficiency and resource utilization, we leverage FP8 quantization and KV caching techniques. Our distributed training leverages a Megatron-based framework, enabling efficient processing of ultra-long user behavior sequences and scalable model training. These infrastructure optimizations resulted in a $\mathbf{57\%}$ improvement in inference speed.
\paragraph{Implementation Details.} We initially used \textbf{Qwen3-14B (Qwen3)}~\citep{yang2025qwen3} as the base model for RecGPT training and dataset accumulation. For user interest mining and item tag prediction tasks, we adapt to the lightweight deployment requirements of online services by training an integrated model based on \textbf{TBStars-MoE-42B-A3.5B (TBStars)} using high-quality training data from the Qwen3 training process. This approach activates only 3.5B parameters per inference request, enabling efficient online service. For explanation generation tasks, we continue using the Qwen3-14B model for both training and inference to ensure high quality and accuracy of the generated results.
\subsection{Online A/B Test}
\label{sec:Online_AB_Test}
\paragraph{Evaluation Metrics.}
We evaluate our online performance across the following dimensions:

\noindent
\textbf{(1) User Experience:}
\begin{itemize}[topsep=0pt, partopsep=0pt]
\item \textbf{Dwell Time (DT):} The average time users spend on the recommended items.
\item \textbf{Exposure Item Category Diversity (EICD):} The diversity of item categories exposed to users.
\item \textbf{Clicked Item Category Diversity (CICD):} The diversity of item categories that users click on.
\end{itemize}

\noindent
\textbf{(2) Platform Benefits:}
\begin{itemize}[topsep=0pt, partopsep=0pt]
\item \textbf{Item Page Views (IPV):} The number of times item pages are viewed from recommendations.
\item \textbf{Click-Through Rate (CTR):} The ratio of clicks to impressions for recommended items.
\item \textbf{Daily Click Active Users (DCAU):} The number of unique users who perform at least one click action on recommended items daily.
\item \textbf{Add-To-Cart (ATC):} The number of items added to the cart from recommendations.
\end{itemize}

The online A/B test results based on TBStars are presented in Table~\ref{tab:ab}, from which we can observe the following key findings:

$\blacklozenge$\;
\textbf{From the user experience perspective}, RecGPT significantly improves user dwell time (DT) and product category diversity (EICD and CICD) by 4.82\%, 0.11\%, and 6.96\%, respectively, by leveraging LLMs' world knowledge and reasoning capabilities to capture users' diverse interest preferences beyond traditional interaction-based methods. Our approach harnesses semantic understanding to infer latent user interests and identify subtle connections between user actions and underlying preferences, enabling recommendations across broader categories while maintaining relevance. The substantial improvement in category diversity demonstrates successful mitigation of the filter bubble effect through uncovering latent preferences, while increased dwell time indicates enhanced user engagement through more serendipitous yet relevant recommendations, ultimately improving user satisfaction and platform experience.

$\blacklozenge$\;
\textbf{From the platform perspective}, RecGPT demonstrates substantial improvements across key engagement metrics. The 9.47\% increase in IPV reflects enhanced user engagement depth, indicating that users are exploring more products per session due to the system's ability to surface genuinely interesting and relevant items that capture their diverse preferences. The 6.33\% boost in CTR demonstrates improved recommendation precision, as users are more likely to click on items that align with their interests captured through our LLM-powered interest modeling and item tag prediction, reducing wasted impressions and improving content relevance. The 3.72\% rise in DCAU signifies improved user retention and platform stickiness, showing that more users are motivated to actively engage with recommendations on a daily basis rather than passively browsing.

$\blacklozenge$\;
\textbf{From the merchant perspective}, RecGPT effectively mitigates the Matthew effect by promoting fairer exposure distribution across merchants of varying scales and popularity levels. As illustrated in the top of Figure~\ref{fig:abs}, our approach demonstrates more uniform CTR performance across different item popularity groups compared to the baseline system. While the baseline system exhibits disproportionate exposure allocation toward high-popularity items, leading to concentration bias that limits competitive opportunities for less popular merchants, RecGPT achieves consistently higher and more stable click-through rates across different popularity groups. This indicates that less popular items receive meaningful exposure opportunities without sacrificing overall performance. Furthermore, as shown in the bottom of Figure~\ref{fig:abs}, the Page View Rate (PVR) distribution reveals that RecGPT effectively flattens the long-tail distribution, providing increased visibility for merchants with lower-popularity items. This redistribution creates more equitable market opportunities, enabling smaller merchants to compete more effectively while maintaining the platform's overall engagement quality, fostering a healthier and more sustainable marketplace ecosystem.

These comprehensive improvements demonstrate that RecGPT successfully creates a \textbf{win-win-win outcome for all ecosystem stakeholders}. By mitigating both filter bubbles for users and the Matthew effect for merchants, our approach enhances user satisfaction through diverse discovery experiences while ensuring fairer market opportunities for merchants of all scales, ultimately strengthening platform health through increased engagement and transaction volume. This multi-stakeholder value creation establishes a virtuous feedback loop where improved user experiences drive higher retention and activity, generating richer behavioral data that enables more precise recommendations, which in turn boost merchant performance and platform growth, creating sustainable incentives for continued ecosystem optimization and long-term competitive advantage.

\begin{table}[t]
    \centering
    \caption{The performance improvement of RecGPT compared to the baseline in the online A/B test conducted from June 17 to June 20, 2025.}
    \label{tab:ab}
    \begin{tabular}{*{9}{c}}
        \toprule
        \textbf{Scenario} & \multicolumn{8}{c}{\textbf{Metrics (Improvement)}} \\
        \midrule
        \multirow{2}{*}{Guess} & DT & EICD & CICD & IPV & CTR & DCAU & ATC \\
        & +4.82\% & +0.11\% & +6.96\% & +9.47\% & +6.33\% & +3.72\% & +3.91\% \\
        \bottomrule
    \end{tabular}
\end{table}

\subsection{Human vs. LLM-as-a-Judge}
\label{sec:LLM_as_a_Judge_Evaluation}
\paragraph{Evaluation Setup.}
To validate the effectiveness of LLM-as-a-Judge methods for recommendation generation tasks, we conduct comprehensive evaluations across three tasks: \textit{User Interest Mining}, \textit{Item Tag Prediction}, and \textit{Recommendation Explanation Generation}. We employ Qwen3 as our base Judge model, referred to as Qwen3-Judge, and enhance its performance through Supervised Fine-Tuning (SFT) on collected human judgment data, referred to as Qwen3-Judge-SFT.
For evaluation standards, each generation output is assessed across multiple criteria using either binary classification (where only ``Yes'' responses indicate passing for that criterion) or multi-level assessment (where only ``Excellent'' and ``Good'' ratings constitute passing for that criterion). Each recommendation generation result is considered qualified only when it passes all evaluation criteria simultaneously.

\paragraph{Evaluation Metrics.}
We utilize Accuracy (ACC), Precision, Recall, and F1 Score as our evaluation metrics to quantify the agreement between the \textit{LLM-as-a-Judge} and \textit{Human Annotators}. Higher metric scores indicate stronger alignment between the two.

\begin{table}[h]
\centering
\caption{Performance comparison between LLM-based Judge models and human expert evaluations across three recommendation generation tasks. Qwen3-Judge-Base represents the original LLM judge model, while Qwen3-Judge-SFT denotes the fine-tuned version trained on human judgment data. The best results are highlighted in \textbf{bold}.}
\label{tab:human_vs_judge}
\begin{tabular}{*{7}{c}}
\toprule
\textbf{Task} & \textbf{Judge Model} & \textbf{ACC} & \textbf{Precision} & \textbf{Recall} & \textbf{F1} \\
\midrule
\multirow{2}{*}{\makecell{User Interest Mining}} & Qwen3-Judge-Base          & 0.6777 & 0.6742   & \textbf{0.9777} & 0.7968 \\
                              & Qwen3-Judge-SFT  &     \textbf{0.7689}  &    \textbf{0.7996} &    0.8575 &   \textbf{0.8275}      \\ 
\midrule
\multirow{2}{*}{\makecell{Item Tag Prediciton}} & Qwen3-Judge-Base          & 0.8741 & 0.9310   & 0.9196 & 0.9253 \\
                              & Qwen3-Judge-SFT      & \textbf{0.9308} & \textbf{0.9714}   & \textbf{0.9463} & \textbf{0.9587} \\ 
\midrule
\multirow{2}{*}{\makecell{Explanation Generation}} & Qwen3-Judge-Base          & 0.5677  & 0.8753   & 0.5677 & 0.6657      \\
                              & Qwen3-Judge-SFT      & \textbf{0.8976} & \textbf{0.9067}   & \textbf{0.8976} & \textbf{0.9016}      \\
\bottomrule
\end{tabular}
\end{table}

\paragraph{Experimental Results.}
The experimental results are shown in Table~\ref{tab:human_vs_judge}, from which we can draw the following conclusions: 

$\blacklozenge$\;
The baseline Qwen3-Judge-Base model shows varying performance across different tasks, achieving 87.41\% accuracy for item tag prediction, 67.77\% for user interest mining, and 56.77\% for recommendation explanation generation. This performance hierarchy reflects the inherent evaluation complexity of each task, with item tag prediction involving relatively objective assessment criteria, while explanation generation requires sophisticated evaluation of content quality, relevance, and factuality. These results demonstrate that vanilla LLMs lack sufficient capability to serve as effective judges for domain-specific recommendation tasks.

$\blacklozenge$\;
The task-aligned Qwen3-SFT-Judge model significantly outperforms the baseline across different tasks and most metrics. Notable accuracy improvements include explanation generation (56.77\% to 89.76\%), user interest mining (67.77\% to 76.89\%), and item tag prediction (87.41\% to 93.08\%). Similar enhancement patterns are observed in precision, recall, and F1 scores. These comprehensive improvements demonstrate that supervised fine-tuning with human judgment data effectively bridges the alignment gap between automated evaluation and human assessment standards, enabling reliable automated evaluation across diverse recommendation scenarios.

These results demonstrate that LLMs can effectively serve as automated judges for recommendation generation tasks by leveraging their powerful human-like reasoning capabilities. Through alignment with task-specific human judgment data, our fine-tuned judge models achieve sufficient accuracy to replace costly and time-consuming manual evaluation processes. Traditional human evaluation, while providing high-quality assessments, suffers from scalability limitations, extended evaluation cycles, and high operational costs that are incompatible with the rapid development demands of modern enterprises. In contrast, our automated LLM-based evaluation framework enables efficient quality assessment at scale, significantly accelerating the iteration cycle for industrial development while maintaining evaluation reliability, thus providing a practical solution for continuous model improvement and deployment in production environments.

\subsection{Case Studies}
\label{sec:Case_Studies}
\begin{figure}[h] 
    \centering
    \includegraphics[width=0.9\linewidth]{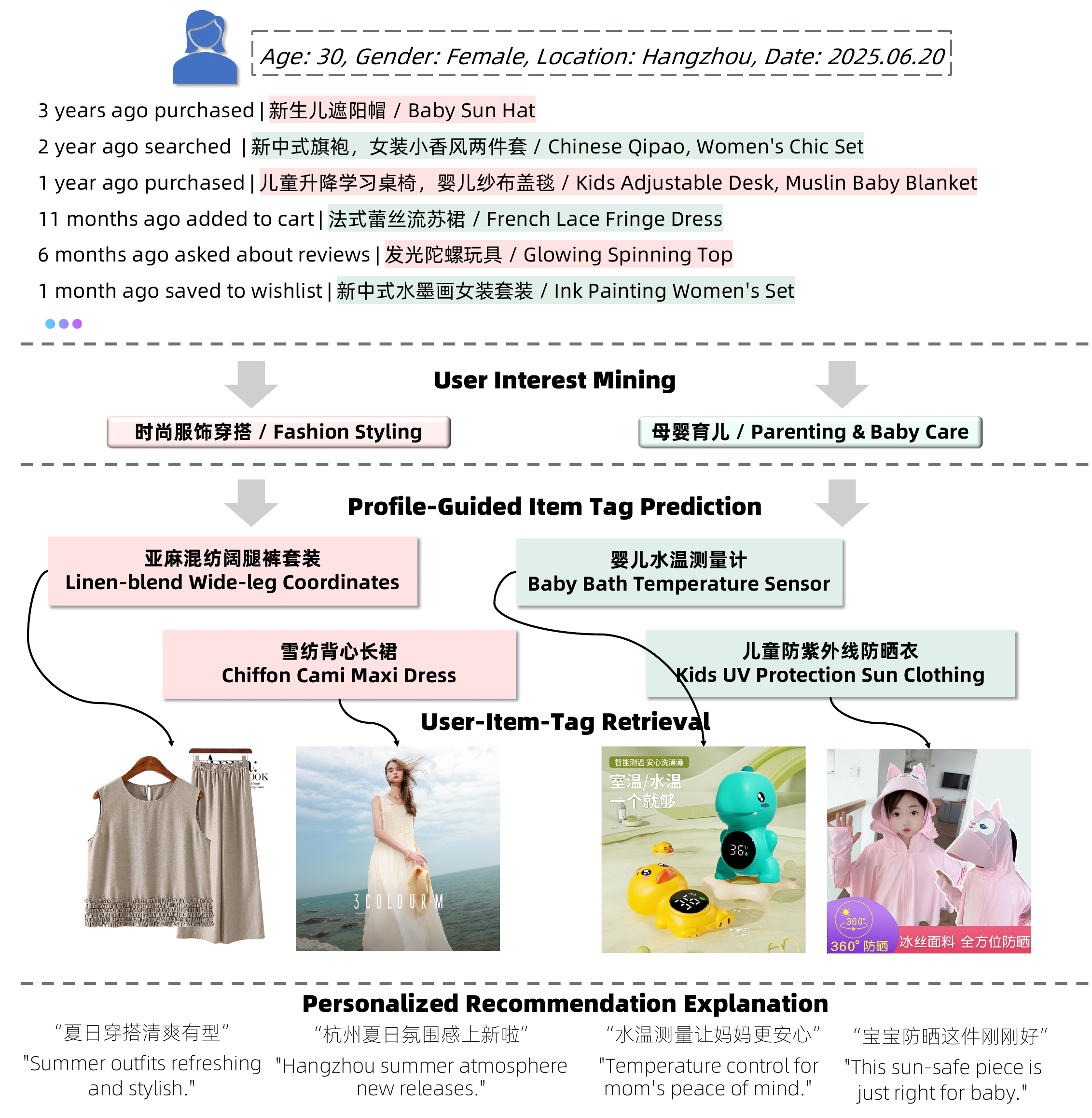}
    \captionsetup{justification=centering}
    \caption{Case Studies of RecGPT in the \textit{Taobao} App's \textit{Guess What You Like} scenario.}
    \label{fig:case_study}
\end{figure}

Figure~\ref{fig:case_study} illustrates the comprehensive workflow of RecGPT through a representative user case, demonstrating its effectiveness in interpreting complex behavioral patterns and generating contextually relevant recommendations tailored to user interests. The example features a \textit{30-year-old female user from Hangzhou} whose extensive \textit{three-year behavioral history} includes diverse purchasing, searching, and browsing activities, indicating distinct preferences for \textit{traditional Chinese fashion aesthetics} and \textbf{modern parenting needs}.

Through systematic analysis of the user's historical activities, such as searches for qipao dresses, women's ink-painting clothing sets, baby sun hats, children's adjustable desks, and glowing spinning toys, the \textbf{User Interest Mining} module identifies two primary areas of interest: \textit{``Fashion styling''} and \textit{``Parenting and baby care''}. These referred interests reflect the system’s ability to detect meaningful thematic patterns within seemingly unrelated behavioral data. Subsequently, the \textbf{Item Tag Prediction} component translates these broad interest categories into specific product-related tags, such as \textit{``Linen-blend Wide-leg Coordinates''} and \textit{``Baby Bath Temperature Sensor''}. These tags effectively capture her preference for stylish yet comfortable fashion and her practical concern for child safety.

The \textbf{User-Item-Tag Retrieval} framework utilizes these tags to select relevant products matching her varied interests. The \textbf{Personalized Recommendation Explanation} module then generates personalized rationales, clearly linking the recommended items to her behavioral history. For example, explanations such as \textit{``Hangzhou summer atmosphere new releases''} seamlessly incorporate her geographic context and seasonal fashion preferences, while \textit{``Temperature control for mom's peace of mind''} directly addresses her emphasis on infant safety. Further context-specific explanations like \textit{``Summer outfits refreshing and stylish''} and \textit{``This sun-safe piece is just right for baby''} resonate with her simultaneous interest in personal style and child protection, completing a \textbf{sophisticated closed-loop system} that effectively translates behavioral insights into meaningful recommendations.

This practical case underscores RecGPT’s core strength: employing task-specific large language models aligned with extensive world knowledge and logical reasoning to reveal users' hidden and diverse interests while maintaining relevance. Unlike traditional collaborative filtering, which relies only on user interactions, RecGPT's LLM-driven approach semantically interprets behaviors, uncovering implicit connections, such as associating traditional fashion interests with cultural identity and linking parenting concerns with safety awareness. This knowledge-driven approach expands recommendation possibilities beyond past interactions, ensuring recommendations remain diverse yet personally meaningful and contextually precise.

\subsection{User Experience Investigation}
\label{sec:User_Experience_Investigation}
\newcommand{\Single}[2]{
  \(\bigcirc\)~#1\quad #2}
\newcommand{\Multi}[2]{
  \(\square\)~#1\quad #2}

\paragraph{Objective}
To systematically validate the effectiveness of the RecGPT in improving recommendation quality, we conduct a comprehensive user study focusing on two critical dimensions:
\begin{itemize}[topsep=0em, itemsep=0.5em]
    \item \textbf{Diversity Assessment:} We evaluate whether RecGPT significantly enhances recommendation diversity by reducing repetition of items with similar brands, categories, or attributes, thereby providing users with richer and more varied choices.
    \item \textbf{User Perception Assessment:} We quantitatively measure improvements in user-perceived recommendation quality through structured feedback collection, with particular emphasis on redundancy perception (e.g., ``Do you feel the recommendations are repetitive?'').
\end{itemize}
\paragraph{Implementation Details}
\begin{itemize}[topsep=0em, itemsep=0.5em]
    \item \textbf{Participant Selection:} We randomly select 500 active users to ensure comprehensive coverage across different demographics, including varied age groups, genders, and interest profiles.
    \item \textbf{Experimental Setup:}
    \begin{itemize}
        \item \textbf{Control Group}: Users receive recommendations generated by the baseline algorithm.
        \item \textbf{Treatment Group}: Users receive recommendations from the RecGPT-enhanced system.
    \end{itemize}
    \item \textbf{Evaluation Methodology:}
    \begin{enumerate}
        \item We use a three-evaluator consensus mechanism where only unanimous decisions are counted as valid responses, ensuring high reliability and minimizing subjective bias.
        \item The evaluation follows a structured three-step process:
    \begin{itemize}
        \item \textbf{Historical Review:} Evaluators examine each user's complete behavioral history (purchases, clicks, interactions, exposures) in chronological order to understand authentic user preferences and browsing patterns.
        \item \textbf{Recommendation Analysis:} Evaluators review the system-generated recommendation lists for both control and treatment groups.
        \item \textbf{Redundancy Assessment:} From the user's perspective, evaluators determine whether obvious redundancy exists in the recommendations and, when present, identify the primary sources of repetition (e.g., dominant product categories, repeated brand patterns, or similar attribute clusters).
    \end{itemize}
    \end{enumerate}
    \item \textbf{Data Analysis:} We conduct comparative analysis of redundancy scores and user satisfaction metrics between treatment and control groups to comprehensively assess RecGPT's impact on recommendation diversity and overall business performance.
\end{itemize}

\paragraph{Questionnaire: Perceived Duplication on the Page}
\noindent
\begin{tcolorbox}[colback=gray!8,colframe=gray!40,
                  boxrule=0.6pt,arc=1mm,
                  enlarge left by=0mm,
                  left=3mm,right=3mm,top=2mm,bottom=2mm,breakable]

\begin{enumerate}[label=\arabic*.]
\item While browsing the current page, do you perceive any duplication?
  \begin{itemize}[leftmargin=8mm,label={},labelsep=0pt]
    \item \Single{A}{Yes, I feel it is duplicated.}
    \item \Single{B}{No, I do not feel any duplication.}
    \item \Single{C}{Skip (layout issue, etc.)}
  \end{itemize}

\item (Multiple choice) If you do perceive duplication, where does it mainly come from?
  \begin{itemize}[leftmargin=8mm,label={},labelsep=0pt]
    \item \Multi{A}{The current main page}
    \item \Multi{B}{The left-hand list}
    \item \Multi{C}{Other, please specify}
  \end{itemize}

\item (Multiple choice) If the duplication comes from the \textbf{current page}, what are the main sources?
  \begin{itemize}[leftmargin=8mm,label={},labelsep=0pt]
    \item \Multi{A}{Too many SKUs with similar specs / prices}
    \item \Multi{B}{Too many variations under the same colour}
    \item \Multi{C}{Overall page looks repetitive}
  \end{itemize}

\item (Multiple choice) If the duplication comes from the \textbf{current page}, which types of products are involved?
  \begin{itemize}[leftmargin=8mm,label={},labelsep=0pt]
    \item \Multi{A}{Exactly the same style}
    \item \Multi{B}{Similar style}
    \item \Multi{C}{Same series}
    \item \Multi{D}{Others (e.g.\ same colour tone, design style)}
  \end{itemize}

\item (Multiple choice) If the duplication comes from the \textbf{left-hand list}, what are the main sources?
  \begin{itemize}[leftmargin=8mm,label={},labelsep=0pt]
    \item \Multi{A}{Repeated items with similar specs}
    \item \Multi{B}{Repeated items under the main list}
    \item \Multi{C}{Repeated tags within the same series}
    \item \Multi{D}{Overlap with the hero image items}
    \item \Multi{E}{Other, please specify}
  \end{itemize}

\item (Multiple choice) If the duplication comes from the \textbf{left-hand list}, which types of products are involved?
  \begin{itemize}[leftmargin=8mm,label={},labelsep=0pt]
    \item \Multi{A}{Exactly the same style}
    \item \Multi{B}{Similar style}
    \item \Multi{C}{Same series}
    \item \Multi{D}{Others (e.g.\ same colour tone, design style)}
  \end{itemize}

\item Please describe any other reasons why you feel the current page is repetitive.\\[1ex]
\end{enumerate}

\end{tcolorbox}

\paragraph{Experimental Results}
Experimental results demonstrate that RecGPT effectively reduces recommendation redundancy across multiple evaluation metrics. Human evaluators identified fewer repetitive items in the RecGPT system, with the repetition rate decreasing from 37.1\% to 36.2\% compared to the baseline. This improvement is most pronounced within the top 4 recommendation slots, where similar product clustering decreased substantially from 27.7\% to 25.3\%, indicating that RecGPT successfully diversifies recommendations in the positions where users focus their attention most.

A notable finding emerged when analyzing advertisement influence on perceived redundancy. The treatment group exhibited more balanced ad distribution patterns, and when ad cards were excluded from the analysis, the reduction in perceived repetition became substantially more pronounced. Specifically, the redundancy improvement nearly doubled from 0.88\% (with ads included) to 1.57\% (without ads), with this effect being most evident within the top 8 recommendation positions.

These findings indicate that RecGPT demonstrates clear advantages in enhancing recommendation diversity and reducing user-perceived repetition. The benefits become particularly apparent when controlling for advertisement interference, suggesting that the model's core recommendation capabilities effectively address content homogenization challenges in practical deployment scenarios.


\section{Conclusion, Limitations, and Future Directions}
\label{sec:Conclusion}
In this paper, we propose RecGPT, a novel recommender system framework that leverages the world knowledge and logical reasoning capabilities of large language models to achieve intent-centered personalized recommendations. RecGPT conducts generative user profiling analysis on users' lifelong multi-behavior sequences and infers users' potential interest distribution through item tag prediction. Additionally, RecGPT enhances system transparency and user experience by generating personalized recommendation explanations.
To align large language models with recommendation domain knowledge, we employ a progressive approach spanning from distillation-based pre-alignment using strong reasoning language models to self-training model evolution. We also transition from expert supervision to an automated Human-LLM cooperative evaluation system, significantly improving both the cost-effectiveness and efficiency of model optimization. Through comprehensive online experiments conducted on Taobao, a real-world e-commerce platform, we validate RecGPT's effectiveness across user experience, commercial conversion, and platform health metrics, demonstrating mutual benefits for users, merchants, and the platform ecosystem.

Although RecGPT has demonstrated promising performance in A/B tests, there are still some limitations and areas for improvement:

$\blacklozenge$\;
\textbf{Modeling Ultra-Long User Sequences}: Handling ultra-long user behavior sequences presents significant challenges for our current model. First, \textbf{\textit{the computational burden is substantial}}, as model training and inference become prohibitively expensive when processing extensive user histories, with approximately 2\% of sequences still exceeding our 128K token limit. Second, \textbf{\textit{maintaining accuracy across such lengthy sequences proves difficult}}, as the model may inadvertently focus on irrelevant noise within user behaviors rather than meaningful interest patterns, resulting in biased user understanding. To address these limitations, we plan to explore advanced sequence modeling techniques specifically designed for LLMs, emphasizing improved \textbf{\textit{Context Engineering}} that can dynamically optimize long-term and short-term memory management, context selection, and information compression for user behavior sequences.

$\blacklozenge$\;
\textbf{Multi-objective Joint Learning with Reinforcement Learning}: Currently, RecGPT relies on supervised learning with periodic model updates, facing two key limitations. First, this static training approach struggles to adapt effectively to continuously evolving user preferences and product characteristics in real-world e-commerce environments. Second, different generation tasks are trained separately without achieving ideal joint optimization, despite their potential for mutual reinforcement as they collectively serve the final recommendation goal. To address these challenges, we plan to develop \textbf{\textit{Reinforcement Learning (RL)}}-based multi-objective joint optimization that utilizes online user feedback data as unified optimization signals. For implementation, we will leverage ROLL~\citep{wang2025reinforcement}, a scalable library designed for large-scale reinforcement learning optimization. This approach will enable joint training across all generation tasks while simultaneously optimizing multiple objectives such as user engagement, conversion rates, and long-term platform health, leading to improved model adaptation through better utilization of real-world user interactions.

$\blacklozenge$\;
\textbf{End-to-End LLM-as-a-Judge Judge System}: Existing RecGPT evaluation frameworks primarily focus on individual task quality assessment, necessitating separate training data for different evaluation dimensions. This approach leads to a fragmented evaluation process that lacks comprehensive contextual understanding and fails to holistically evaluate multiple aspects simultaneously. To address these limitations, we plan to develop an end-to-end LLM-as-a-Judge system incorporating \textbf{\textit{Reinforcement Learning from Human Feedback (RLHF)}}~\citep{kirk2023understanding,casper2023open,lee2023rlaif,kaufmann2024survey} to train LLM-Judge with human feedback for integrated multi-task assessments. Additionally, we will explore inference-scaling generative reward models~\citep{liu2025inference,guo2025reward,chen2025reasongrm}, allowing dynamic allocation of computational resource during inference to enhance evaluation effectiveness and achieve more nuanced pipeline assessments.

How to effectively leverage large language models in real-world industrial recommender systems has attracted significant research attention since the emergence of ChatGPT. As one of the early successful attempts to fully deploy large language models in real applications, RecGPT serves billions of users and products, demonstrating the tremendous potential of LLMs-for-RS. As large language models continue to evolve and application scenarios expand, we will continue exploring how to better utilize their powerful reasoning and generation capabilities to enhance the intelligence level of recommender systems, conducting meaningful and practical research and practices.
\newpage


\addcontentsline{toc}{section}{References}
\bibliographystyle{abbrvnat}
\nobibliography*
\bibliography{bibtex}

\clearpage

\appendix
\section*{Appendix}
\label{sec:Appendix}
\section{Contributions}

\begin{multicols}{2}
\noindent
\textbf{Core Contributors} \\
Chao Yi \\
Dian Chen \\
Gaoyang Guo \\
$\text{Jiakai Tang}^{\dagger}$ \\
Jian Wu \\
Jing Yu \\
Mao Zhang \\
$\text{Sunhao Dai}^{\dagger}$ \\
Wen Chen \\
Wenjun Yang \\
Yuning Jiang \\
Zhujin Gao \\

\noindent
\textbf{Contributors} \\
Bo Zheng \\ 
Chi Li \\
Dimin Wang \\
Dixuan Wang \\
Fan Li \\
Fan Zhang \\
Haibin Chen \\
Haozhuang Liu \\
Jialin Zhu \\
Jiamang Wang \\
Jiawei Wu \\
Jin Cui \\
Ju Huang \\
Kai Zhang \\
Kan Liu \\
Lang Tian \\
Liang Rao \\
Longbin Li \\
Lulu Zhao \\
Na He \\
Peiyang Wang \\
Qiqi Huang \\
Tao Luo \\
Wenbo Su \\
Xiaoxiao He \\
Xin Tong \\
$\text{Xu Chen}^{\dagger}$ \\
Xunke Xi \\
Yang Li \\
Yaxuan Wu \\
Yeqiu Yang \\
Yi Hu \\
Yinnan Song \\
Yuchen Li \\
Yujie Luo \\
Yujin Yuan \\
Yuliang Yan \\
Zhengyang Wang \\
Zhibo Xiao \\
Zhixin Ma \\
Zile Zhou \\
Ziqi Zhang \\
\end{multicols}

\noindent
$\dagger$ Renmin University of China

\noindent
The listing of authors is in alphabetical order based on their first names.

\newpage

\section{Prompts}
\begin{tcolorbox}[promptbox=interest_colframe/interest_colback, title={User Interest Mining Prompt Template},label=box:app-interest-mining,breakable]
\colorbox{green!20}{\textbf{\# Role}}\\
You are a shopping guide for an e-commerce platform. Based on users' behavioral history, you need to accurately and comprehensively analyze their potential interests and preferences.\\
\colorbox{blue!20}{\textbf{\# Input}}\\
\textbf{User Profile}: \textcolor{BlueViolet}{\textbf{\{Generated User Attributes\}}}

\textbf{User Behavioral Information}:\\
Click Behavior Sequence: \textcolor{BlueViolet}{\textbf{\{Click Behavior Sequence\}}} | 
Purchase Behavior Sequence: \textcolor{BlueViolet}{\textbf{\{Purchase Behavior Sequence\}}} |
Search Behavior Sequence: \textcolor{BlueViolet}{\textbf{\{Search Behavior Sequence\}}} |
Extra Information: \textcolor{BlueViolet}{\textbf{\{Extra Information\}}}\\
\colorbox{red!20}{\textbf{\# Mandatory Requirements}}\\
\textbf{**Interest Mining Principles(\correct)**} \\
1.Differentiate long-term interests from short-term usage scenarios. \\
2.Mine high-confidence interests based on diverse behaviors. \\
3.Validate interest credibility through temporal distribution patterns and eliminate non-sustained behaviors.\\
\textbf{**Interest Mining Requirements (\correct)**} \\
1.Cross-reference gender, age, and other attributes to exclude gift-related scenarios inconsistent with user characteristics. \\
2.Categorize results into listed interests and extended interests. 

\textbf{**Quantity Requirements (\correct)**} \\
Reason over 10 interests.

\textbf{**Task Constraints (\wrong)**} \\
1.Avoid selecting interest categories unrelated to user behavior. \\
2.Avoid exclude daily consumables.\\
3.When evaluating, do not rely on single actions or short-term concentrated purchases.\\
\colorbox{orange!20}{\textbf{\# Preset Interest List}}\\
\textcolor{Maroon}{\textbf{\{Matched Interest Pool\}}}\\
\colorbox{yellow!20}{\textbf{\# Output Format}}
\begin{verbatim}
[
  {
    "ID": "matched interest_01",
    "Interest": "xxx",
    "Stage": "yyy",
    "Reason": "zzz",
  },
  ...
]
\end{verbatim}
\end{tcolorbox}

\clearpage

\begin{tcolorbox}[promptbox=tag_colframe/tag_colback, title={Item Tag Prediction Prompt Template},label=box:app-tag-prediction,breakable]
\colorbox{green!20}{\textbf{\# Role}}\\
You are a professional product recommendation specialist for the Taobao app.\\
\colorbox{blue!20}{\textbf{\# Input}}\\
\textbf{User Attributes}:
\textcolor{BlueViolet}{\textbf{\{Generated User Attributes\}}}\\
\textbf{User Interests}:
\textcolor{BlueViolet}{\textbf{\{Generated User Interests\}}}\\
\textbf{User Behavior Information}\\
Click Behavior Sequence: \textcolor{BlueViolet}{\textbf{\{Click Behavior Sequence\}}} | 
Purchase Behavior Sequence: \textcolor{BlueViolet}{\textbf{\{Purchase Behavior Sequence\}}} |
Search Behavior Sequence: \textcolor{BlueViolet}{\textbf{\{Search Behavior Sequence\}}} |
Extra Information: \textcolor{BlueViolet}{\textbf{\{Extra Information\}}}\\

\colorbox{red!20}{\textbf{\# Mandatory Requirements}}\\
\textbf{Task Requirements}:  \\
\textbf{**Recommendation Principles (\correct)**} \\
1.Combine user profiles and behavior sequences. \\
2.Focus primarily on long-term and recent behaviors. \\
3.Pay attention to current time and seasonal factors. 

\textbf{**Recommendation Requirements (\correct)**} \\
1.Provide specific item descriptions without mentioning brand or model, but avoid overly broad descriptions. \\
2.Associate each item with a user's interest preference. \\
3.Give the recommendation rationale, linking it to the user's relevant attributes and historical behavior. 

\textbf{**Quantity Requirements (\correct)**} \\
Recommend 50 products.

\textbf{**Strict Prohibitions (\wrong)**} \\
1.Items that the user has clicked on, purchased, or searched for within the past month should not be recommended. \\
2.Avoid using broad or vague descriptive terms such as “smart”, “modular”, or “set” in item descriptions.

\colorbox{yellow!20}{\textbf{\# Output Format}}
\begin{verbatim}
[
  {
    "Item Tag": "xxx",
    "Interest": "yyy",
    "Reason": "zzz",
  },
  ...
]
\end{verbatim}
\end{tcolorbox}

\clearpage

\begin{tcolorbox}[promptbox=exp_colframe/exp_colback, title={Recommendation Explanation Generation Prompt Template},label=box:app-exp-generation,breakable]
\colorbox{green!20}{\textbf{\# Role}}\\
Generate personalized recommendation explanations based on user profiles and recommended items. The explanations must satisfy the following requirements.\\
\colorbox{blue!20}{\textbf{\# Input}}\\
\textbf{User Interest}: \textcolor{BlueViolet}{\textbf{\{User Interest\}}}\\
\textbf{Current Date}: \textcolor{BlueViolet}{\textbf{\{Date Information\}}}\\
\textbf{Item Information}: \textcolor{BlueViolet}{\textbf{\{Item Information\}}}\\
\colorbox{orange!20}{\textbf{\# Core Reasoning Steps}}\\
1.\textbf{Context Understanding}: Extract core item characteristics from the item title and extra information.\\
2.\textbf{Explanation Generation}: Synthesize creative and impactful recommendation explanations through analytical interpretation of input information.\\
\colorbox{red!20}{\textbf{\# Mandatory Requirements}}\\
\textbf{**Recommendation Principles (\correct)**}\\
1.\textbf{Length}: 6–10 characters (exclusive of punctuation/spaces)\\
2.\textbf{Style}: Naturally fluent phrasing with concrete descriptions\\
3.\textbf{Expression}: Incorporate humor, wit, and digital-native flair; metaphorical or homophonic techniques are permitted\\
\textbf{**Strict Prohibitions (\wrong)**}\\
1.Fabricating functional benefits, materials, or brand attributes\\
2.Meaningless generic terms (e.g., "practical," "versatile," "elegant")\\
3.Slogan-style clichés (e.g., "essential gadget")\\
4.Exaggerated claims (e.g., "100\% effective")\\
5.Near-duplication of product titles\\
\colorbox{yellow!20}{\textbf{\# Output Format}}
\begin{verbatim}
{
    "Explation": "xxx"
}
\end{verbatim}
\end{tcolorbox}


\section{Implementation Details of Curriculum Learning-based Multi-task Fine-tuning}
\label{sec:Curriculum_Learning_Multi_task_Fine_tuning}

\begin{figure}[t]
\centering
\includegraphics[width=\linewidth]{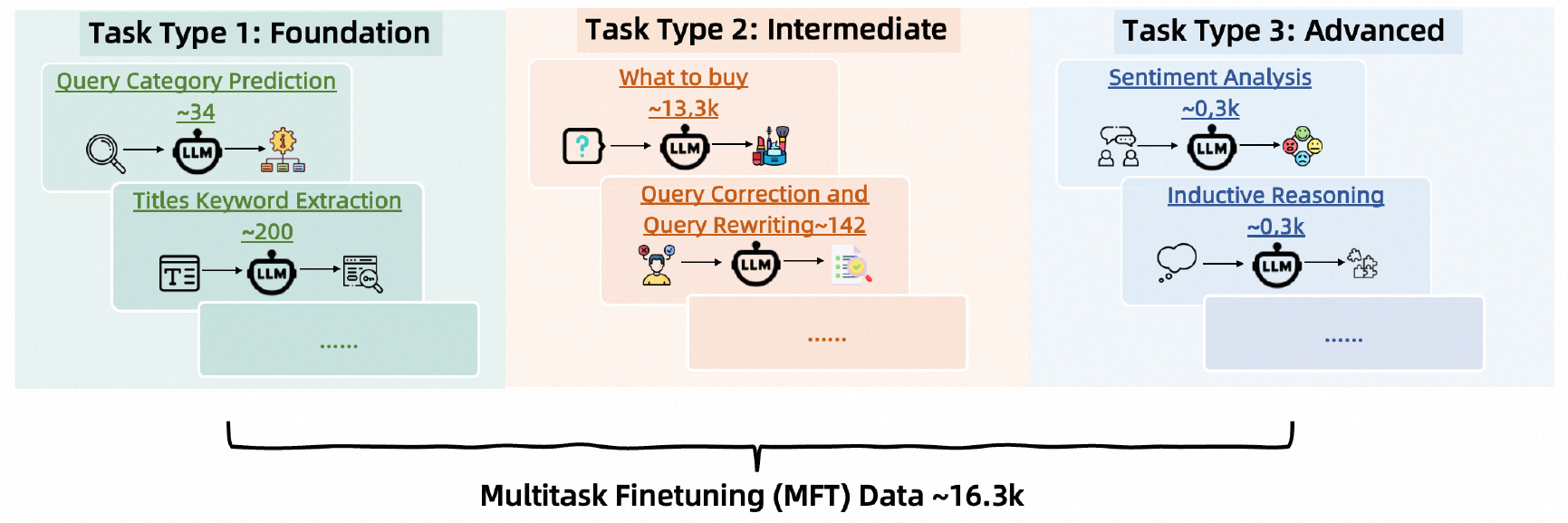}
\caption{Curriculum learning-based multi-task fine-tuning framework with foundation, intermediate, and advanced levels, progressively guiding LLMs to develop capabilities for solving complex e-commerce domain tasks.}
\label{fig:mft}
\end{figure}

Directly applying large language models to user interest mining in e-commerce domains presents significant challenges that limit their effectiveness:
\begin{itemize}[itemsep=0.5em,topsep=0.em]
\item \textbf{Open Interest Space and Uncertainty:} Users' interest landscapes are expansive, and their affinity for previously unseen interests exhibits high uncertainty, making it difficult to effectively introduce new interests to users.
\item \textbf{Task Complexity Disparity:} Industrial-scale LLMs lack deep understanding of the massive, rapidly evolving item corpora on online platforms. Off-the-shelf LLMs cannot effectively capture domain-specific behavioral patterns.
\end{itemize}

To bridge this gap between general LLM capabilities and domain-specific requirements, we propose a \textbf{Curriculum Learning-based Multi-task Fine-tuning (CL-MFT)} approach. This framework guides the model through progressively challenging tasks, beginning with fundamental e-commerce concepts and advancing to sophisticated interest inference and recommendation generation. By following this structured learning progression, the model develops robust domain knowledge while maintaining strong generalization capabilities for complex user behavior modeling and cold-start scenarios.

Our curriculum design (see Figure~\ref{fig:mft}) organizes tasks into three progressive difficulty levels that mirror natural learning progression: \textbf{foundation}, \textbf{intermediate}, and \textbf{advanced}. The complete task list is presented in Table~\ref{tab:mft_samples}.

At the \textbf{foundation level}, we establish core competencies through tasks that directly connect user behavior with purchase intent. Specifically, \textit{Query Category Prediction} teaches the model to map search terms to product categories, building essential connections between user actions and interest signals. \textit{Query-Item Relevance Judgment} develops the ability to evaluate matches between search queries and product titles, uncovering users' attribute-specific preferences. \textit{Key Information Extraction from Item Titles} trains the model to identify crucial product characteristics (e.g., retro style, silent operation), forming the foundation for comprehensive interest profiling.

The \textbf{intermediate level} advances to tasks requiring sophisticated profile integration. For example, The \textit{E-commerce What-to-Buy} task challenges the model to analyze contextual queries (e.g., 0-1 year old baby clothes) and infer broader interest domains (e.g., maternal and infant products) while considering user demographic information. \textit{Query Correction and Query Rewriting} develop refinement capabilities, transforming imprecise searches into targeted intent (e.g., ``cream-style bathroom cabinet'' $\to$ ``cream-colored solid wood bathroom cabinet''), thereby improving interest matching precision.

\begin{table}[h]
\centering
\caption{Curriculum Learning-based Multi-task List. Each task is categorized by type and subtask, with the number of samples used for training.}
\label{tab:mft_samples}
\centering
\begin{tabular}{|c|c|c|}
\hline
\textbf{Task Type} & \textbf{Subtask} & \textbf{Count} \\
\hline
Foundation & Query Category Prediction & 34\\
\hline
Foundation & Query-Item Relevance  & 100 \\
\hline
Foundation & Key Information Extraction from Item Titles & 200 \\
\hline
Foundation & Item Key Points Extraction & 100\\
\hline
 Intermediate & E-commerce \textit{What to Buy} &  13.3k\\
 \hline
 Intermediate & E-commerce Concept Explanation & 200 \\
\hline
Intermediate & Unique Selling Proposition &  200 \\
 \hline
 Intermediate & Query Correction & 44 \\
 \hline
Intermediate &  Query Rewriting & 108 \\
\hline

Adavanced & Recommandation & 300 \\
\hline
Adavanced & Keyword Extraction & 300 \\
\hline
Adavanced & Sentiment Analysis & 300 \\
\hline
Adavanced & Text Classification & 300 \\
\hline
Adavanced & Inductive Reasoning & 300 \\
\hline
Adavanced & Deductive reasoning & 212 \\
\hline
Adavanced & References & 300\\
\hline
\end{tabular}
\end{table}

The \textbf{advanced level} introduces complex reasoning through \textit{Causal Reasoning} and \textit{Inductive Reasoning} tasks that decode behavioral patterns (e.g., ``waterproof headphones'' $\to$ ``outdoor sports affinity''). \textit{Keyword Extraction and Sentiment Analysis} extract nuanced insights from user reviews to validate and refine interest inferences. Finally, \textit{Product Description Generation} task synthesizes all acquired capabilities to produce content that precisely aligns with novel user interests.

This staged training approach introduces low-complexity, diverse tasks first, reducing the model's learning burden and laying the groundwork for subsequent interest mining. It ultimately trains the base LLM to transition from simple matching to comprehensive reasoning for interest discovery.

\end{document}